\documentclass[conf]{new-aiaa}
\usepackage[utf8]{inputenc}

\usepackage{graphicx}
\usepackage{amsmath}
\usepackage[version=4]{mhchem}
\usepackage{siunitx}
\usepackage{longtable,tabularx}
\usepackage{gensymb}
\usepackage{slashbox,multirow}
\usepackage{caption}  
\usepackage{multibib}
\usepackage{float}
\usepackage{caption}
\usepackage{subcaption}
\usepackage{bm}

\title{Assessment of Detached Eddy Simulation and Sliding Mesh Interface in Predicting Tiltrotor Performance
in Helicopter and Airplane Modes}

\author{Feilin Jia\footnote{Research Scientist, Department of Computational Fluid Dynamics.} and John Moore\footnote{Chief Executive Officer, Department of Computational Fluid Dynamics.}}
\affil{FlexCompute, 130 Trapelo Road, Belmont, MA, 02478}
\author{Qiqi Wang\footnote{Associate Professor, Department of Aeronautics and Astronautics, AIAA Associate Fellow.}}
\affil{Massachusetts Institute of Technology, 77 Mass Ave, Cambridge, MA, 02139}

\begin{document}
\maketitle

\begin{abstract}
This paper presents numerical investigation on performance and flow field of the full-scale XV-15 tiltrotor in both helicopter mode (hovering flight and forward flight) and aeroplane propeller mode using Detached Eddy Simulation, in which the movement of the rotor is achieved using a Sliding Mesh Interface. Comparison of our CFD results against experiment data and other CFD results is performed and presented.
\end{abstract}
\section{Nomenclature}
{\renewcommand\arraystretch{1.0}
\noindent\begin{longtable*}{@{}l @{\quad=\quad} l@{}}
$R$  & rotor disk radius\\
$A$  & rotor disk area, $\pi R^2$\\
$c$  & blade chord \\
$c_{ref}$ &    reference blade chord \\
$C_p$& surface pressure coefficient \\
$C_Q$ & rotor torque coefficient, $Q/\rho (\Omega R)^2AR$ \\
$C_T$ & rotor thrust coefficient, $T/\rho (\Omega R)^2A$\\
FoM & figure of merit, FoM=$\frac{C_T^{3/2}}{\sqrt2C_Q}$\\
$T$ & rotor thrust\\
$Q$ & rotor torque\\
$\Omega$ & rotor angular speed\\
$\tau_{wall}$ & wall shear stress
\end{longtable*}}

\section{Introduction}
Over the past decades, tiltrotor has attracted more and more attention because of the emerging demands for a new type of flying vehicle that has capabilities of both vertical take-off/landing (VTOL) and high-speed cruise. After this technology was demonstrated with the Bell XV-3 in 1955 for the first time~\cite{Maisel2000TheFlight}, a joint program was launched by the NASA Ames Research Center and Bell Helicopters to develop a new tiltrotor named XV-15 in the late 1960s and early 1970s. The XV-15 tiltrotor research aircraft took its first flight in 1977, and it has been substantially utilized to support many other tiltrotor research activities since then, e.g. Bell-Boeing V-22 Osprey\cite{Potsdam2004TiltFlight,Narducci2009CFDInteractions} and the AW609\cite{Harris2010InitialTiltrotor}.
Unlike typical helicopter rotors, tiltrotor blades demand a compromised design to operate efficiently in both helicopter and propeller modes, so the blades of XV-15 tiltrotor have high twist and solidity, along with small rotor radius. Because of its complexity in geometry and widely varying operation conditions, many experimental investigations of the rotor performance of the XV-15 tiltrotor have been done over the years. The first experiment of the full-scale XV-15 rotor was employed in the NASA 40-by-80-Foot Wind Tunnel\cite{Anon1971AdvancementResults} in 1970. In 1980, Weiberg et al. \cite{Weiberg1980Wind-TunnelAircraft} used the same facility to measured integrated rotor loads in helicopter, aeroplane, and transition-coordidor modes. However, the measured force and moment included the contribution from the airframe. In 1985, a comprehensive experimental study of the XV-15 tiltrotor in helicopter mode was conducted by Felker et al. \cite{Felker1986PerformanceRotor} at the NASA-Ames Outdoor Aeronautical Research Facility (OARF). In 1997, Light \cite{Light1997ResultsComplex} conducted hover and forward flight tests in the 80-fit by 120-ft Wind Tunnel, but the number of tested operation conditions are relatively limited. To fill the gap, in 2002, Betzina did a series of experiments on hovering flight and propulsive/descending forward flight in helicopter mode, and then an extensive data set was published \cite{Betzina2002RotorMode}. It should be noted that all of the above cited experiments didn't include measurement of surface skin friction, which is very useful to find the transition location and identify regions of reversed flow to facilitate the design of blades. Therefore, Wadcook et al. \cite{Wadcock1999SkinRotor} measured skin friction on a hovering full-scale XV-15 tiltrotor. \par

In recent decades, both computer hardware and numerical algorithms have evolved and advanced dramatically, which makes computational fluid dynamics (CFD) a powerful tool to predict rotor performance. Significant efforts have been performed to develop various numerical methods to improve the accuracy of CFD simulations on tiltrotors compared with the published experimental data. The simplest method was analytical models based on the actuator disk theory and the blade element momentum theory \cite{Johnson1980HelicopterTheory}, but it ignores the blade-vortex interaction and the effect of rotor wake. To overcome the above drawback, several prescribed wake models were proposed \cite{Landgrebe1972WAKEPERFORMANCE.,Kocurek1977PRESCRIBEDANALYSIS.}. However, because these models don't take the effects of tip shape, nonlinear blade twist and flow separation into account rigorously, today they are only used for preliminary performance estimates in some low-fidelity CFD codes. In order to simulate the near blade flow at higher degree of accuracy on limited computing resources, some hybrid solvers were developed, which combined a Navier-Stokes solver in the unsteady viscous flow regions near the blades and a potential flow solver for wake convection in the farfield \cite{Schmitz2009TheLoads,Yang2002RecentFlight}. Due to the emergence of parallel computers, modern high fidelity approaches based on full-domain numerical simulation of the Navier-Stokes equations were gradually employed. Kaul et al. \cite{Kaul2011SkinOVERFLOW2, Kaul2012EffectFlows} and Yoon et al. \cite{Yoon2014SimulationsOverflow} conducted simulations on the hovering XV-15 main rotor blade using OVERFLOW CFD code. Sheng et al. \cite{Sheng2016InvestigationsCodes} used the U\textsuperscript{2}NCLE and Helios CFD solvers to assess the impact of transition models on prediction of the hover figure of merit on the XV-15 blade. Gates \cite{Gates2013AerodynamicComputations} and Garcia et al \cite{Jimenez-Garcia2017TiltrotorValidation} used Helicopter Multi-Block (HMB) CFD solver to perform detailed performance analyses of the hover and propeller modes of the XV-15 blades. However, there hasn't been a comprehensive numerical study on propulsive/descending forward flight helicopter modes of XV-15 tiltrotor.\par
In this paper, the hovering and forward flights of helicopter mode and airplane mode were investigated numerically by our CFD solver: Flow360. In Sec.\ref{section:geometryMesh}, the geometry and the mesh generation methodology of multi-block meshes are introduced. In Sec.\ref{section:numericalMethods}, the numerical methods used in the present study are briefly reviewed. In Sec.\ref{section:testConditions}, the various operating conditions for the hover and forward helicopter modes and airplane mode are listed. In Sec.\ref{section:results}, the numerical results are shown and compared with published experimental data followed by some detailed studies and discussions. Finally, we conclude the paper with some major findings in Sec.\ref{section:conclusions}.

\section{XV-15 Rotor Geometry and Mesh}
\label{section:geometryMesh}
\subsection{XV-15 Rotor Geometry}
The CFD model of XV-15 rotor was generated based on the experiment by Felker \cite{Felker1986PerformanceRotor}. The rotor blade consists of 5 NACA 6-series aerofoil sections, which are reported in Table~\ref{tab:1}.

\begin{table}[H]
\centering
\caption{\label{tab:1}Radial location of the XV-15 rotor blade aerofoils}
\begin{tabularx}{0.8\textwidth} { 
  | >{\centering\arraybackslash}X 
  | >{\centering\arraybackslash}X | }
 \hline
 \textbf{r/R} & \textbf{Aerofoil} \\
 \hline
 0.09  & NACA 64-935   \\
\hline
0.17  & NACA 64-528   \\
\hline
0.51  & NACA 64-118   \\
\hline
0.80  & NACA 64-(1.5)12   \\
\hline
1.00  & NACA 64-208   \\
\hline
\end{tabularx}
\end{table}

\begin{figure}[H]
    \centering
    \includegraphics[width=0.75\textwidth]{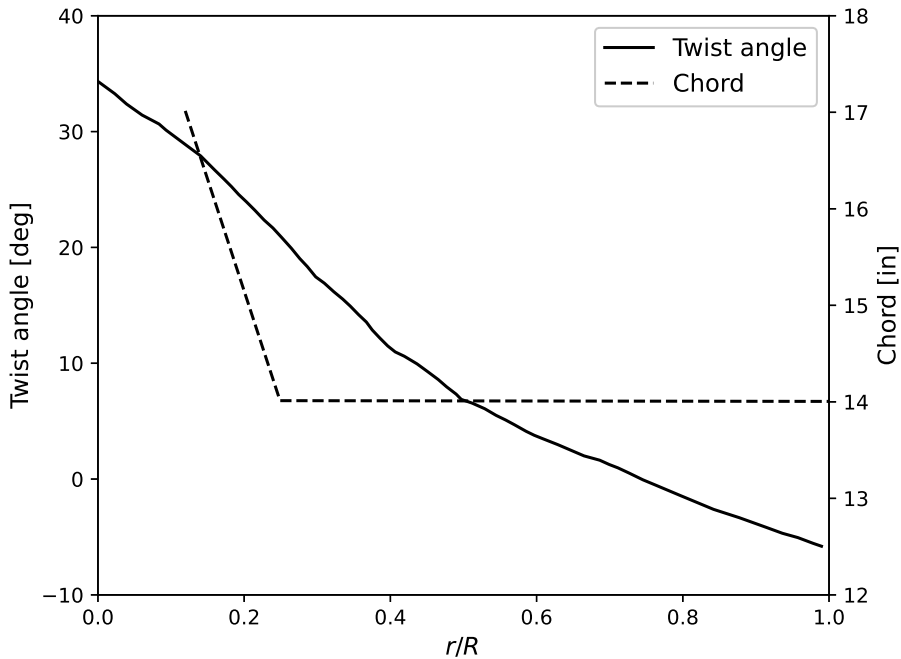}
    \caption{XV-15 rotor blade's twist and chord radial distribution~\cite{Felker1986PerformanceRotor}.}
    \label{fig:1}
\end{figure}

In the present paper, the tool to create the geometry file of the blade is The Engineering Sketch Pad V1.18 \cite{Haimes2013TheGeometry}. The radius of the three-bladed rotor is 150 inches. The inboard aerodynamic section starts at 9.1\% radius with a chord of 16.6 inches, linearly tapering a chord of 14 inches by 25\% radius. The chord keeps a constant 14 inches from 25\% radius to the tip. Each blade has a structural twist angle of $-40.25\degree$ from the root cutoff to the tip. The main geometric characteristics of the XV-15 rotor blades are listed in Table~\ref{tab:2}. Also, a detailed sketch of the XV-15 blade radial twist and chord distributions are shown in Figure~\ref{fig:1}. Different from the conventional helicopter blades, tiltrotor blades are characterised by high twist and solidity, along with a small rotor radius.

\begin{table}[H]
\centering
\caption{\label{tab:2}Geometric properties of the full-scale XV-15 rotor}
\begin{tabularx}{0.8\textwidth} { 
  | >{\centering\arraybackslash}X 
  | >{\centering\arraybackslash}X | }
 \hline
 \textbf{Parameter} & \textbf{Value} \\
 \hline
 Number of blades, $N_b$  & 3   \\
\hline
 Rotor radius, $R$  & 150 inches   \\
\hline
 Reference blade chord, $c_{ref}$  & 14 inches   \\
\hline
 Aspect ratio, $R/c_{ref}$ & 10.71   \\
\hline
 Rotor solidify, $\sigma$ & 0.089   \\
\hline
 Linear twist angle, $\Theta$ & -40.25\degree   \\
\hline
\end{tabularx}
\end{table}

\subsection{XV-15 Rotor Mesh}
\begin{figure}[H]
\begin{subfigure}{.5\textwidth}
  \centering
  \includegraphics[width=.8\linewidth]{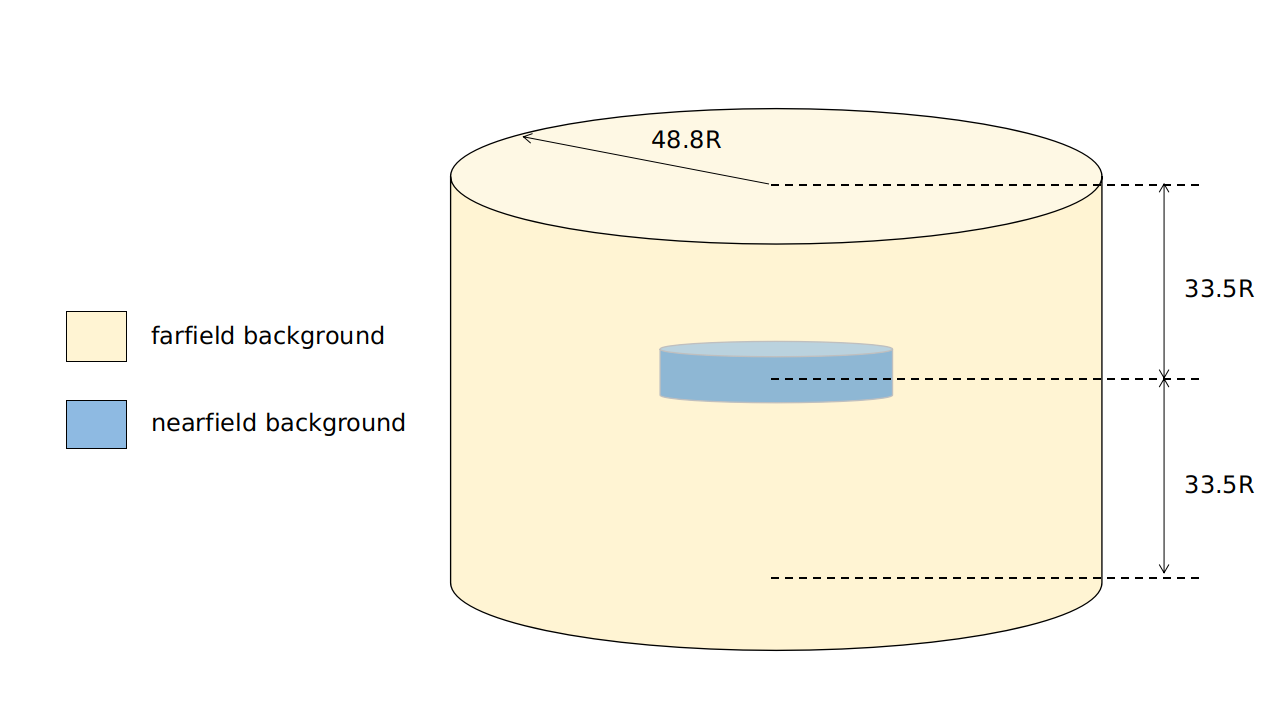}  
  \caption{Dimensions of the farfield background domain}
  \label{fig:farfield_sketch}
\end{subfigure}
\begin{subfigure}{.5\textwidth}
  \centering
  \includegraphics[width=.8\linewidth]{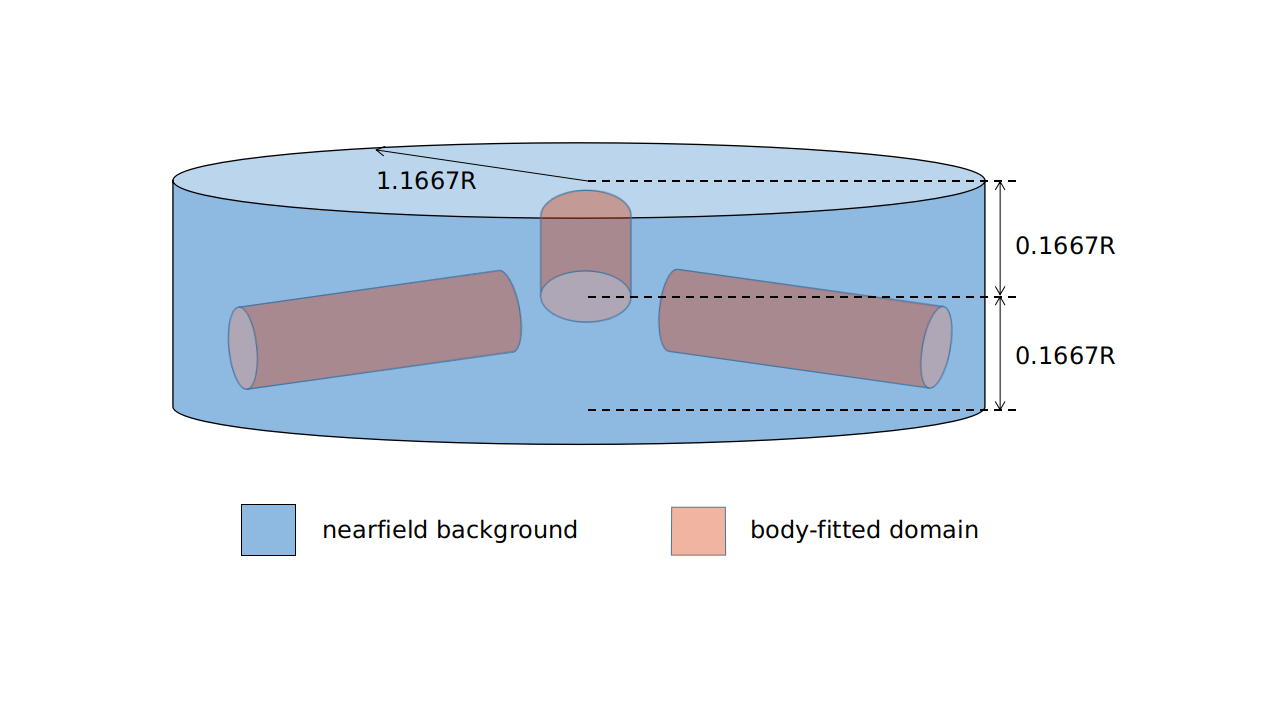}  
  \caption{Dimensions of the nearfield background domain}
  \label{fig:nearfield_sketch}
\end{subfigure}
\newline
\begin{subfigure}{.5\textwidth}
  \centering
  \includegraphics[width=.8\linewidth]{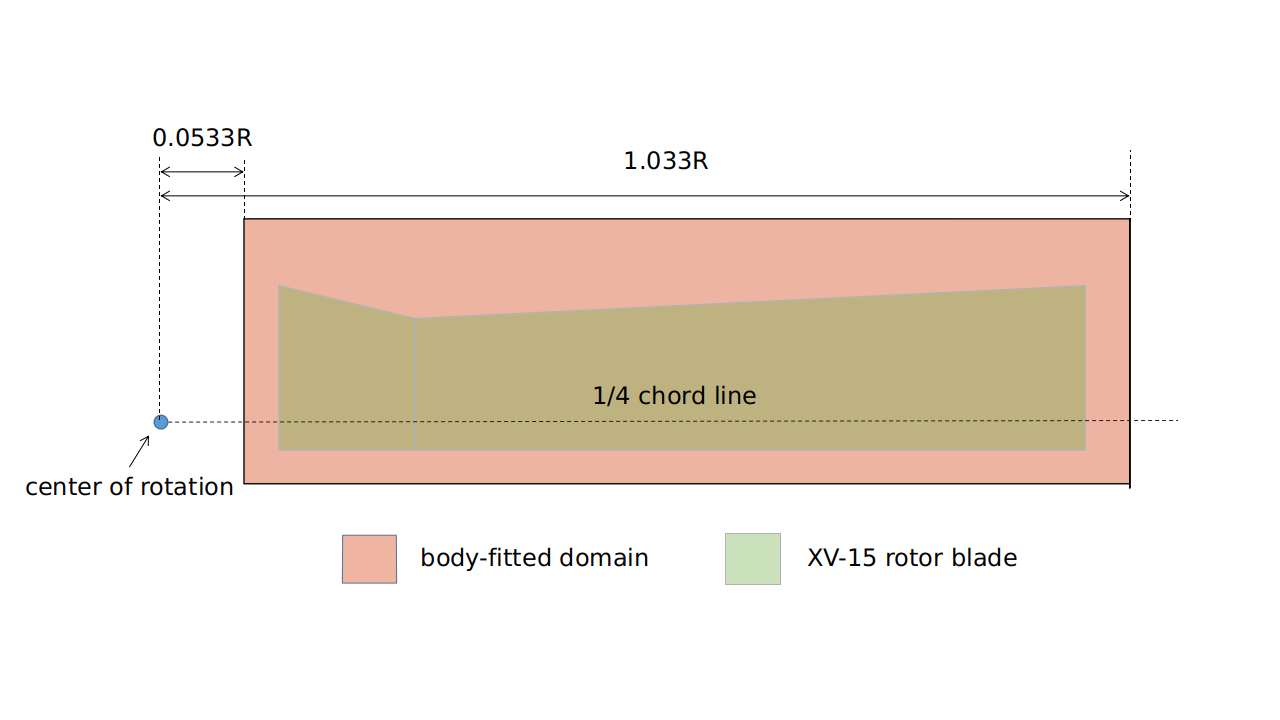}  
  \caption{Dimensions of the body-fitted domain}
  \label{fig:body-fitted_sketch}
\end{subfigure}
\begin{subfigure}{.5\textwidth}
  \centering
  \includegraphics[width=.8\linewidth]{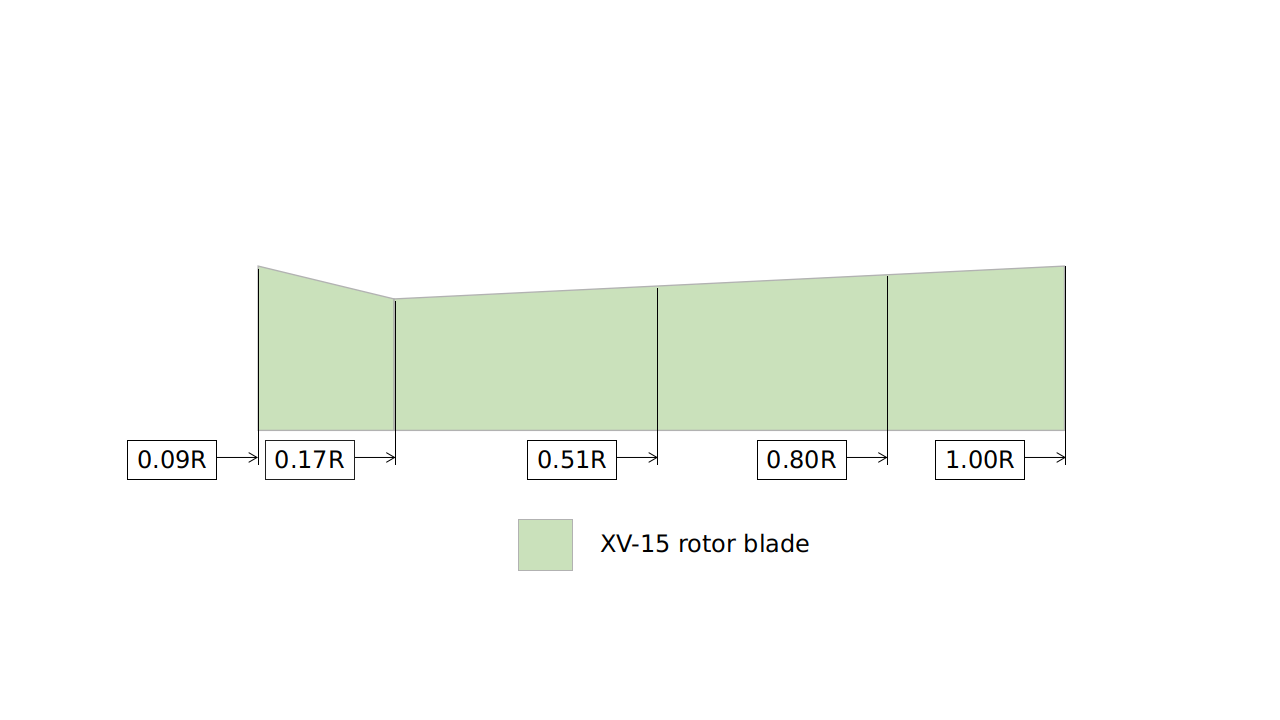}  
  \caption{Planform of the XV-15 rotor blade}
  \label{fig:blade_sketch}
\end{subfigure}
\caption{Sketch of the domains of farfield background, nearfield background and body-fitted for meshing and the Planform of the XV-15 rotor blade.}
\label{fig:multiple_domains_xv15}
\end{figure}

A multi-block unstructured mesh was generated for the aerodynamic study of the XV-15 rotor. The entire domain was splitted into multiple blocks for meshing, shown in Figure~\ref{fig:multiple_domains_xv15}. The farfield block acts as the stationary domain, and the nearfield block, containing the three blades, acts as the rotational domain. For the farfield block, the boundaries were extended to $33.5R$ above and below the rotor disk plane in the axial direction, and to $48.8R$ from the center of domain in the radial direction, which are far enough to assure an independent solution with the boundary conditions employed. For the nearfield block, because CFD simulations of the XV-15 tiltrotor need to be conducted over a range of blade collective angles from $0$ to $18$ degrees, it is more flexible to split the nearfield block into 4 parts: 1 cylindrical off-body mesh (nearfield background) and 3 cylindrical body-fitted meshes containing the blades. It should be noted that, to avoid non-conforming meshes, the nodes on the interfaces between the nearfield background mesh and body-fitted meshes are one-to-one matched. The nearfield background mesh could be reused for all body-fitted meshes with various blade collective angles. It is extended to 0.1667R above and below the rotor plane in the axial direction and to 1.1667R in the radial direction. For the body-fitted meshes, one cylindrical body-fitted mesh was generated first. Its axial direction is along the radial direction of the computational domain. It expanded from 0.1367R to 1.0333R in the radial direction and the radius of its circular face is 0.1024R. Then it was rotated by 120 degree and -120 degree respectively along the axial direction of the computational domain to obtain the other two body-fitted meshes by our mesh transformation tool. Finally, the off-body mesh and the 3 body-fitted meshes were merged together to obtain the nearfield block by our efficient mesh merging tool. All the overlapped nodes from the above meshes are detected and merged into one single node, so the resulting merged nearfield block is a single-block conforming mesh. During the simulation, the inner block rotates along the axial direction of the domain and outer block remains stationary. The solution is communicated through the interface of the above two blocks via interpolation.

\begin{figure}[H]
\begin{subfigure}{.5\textwidth}
  \centering
  \includegraphics[width=.8\linewidth]{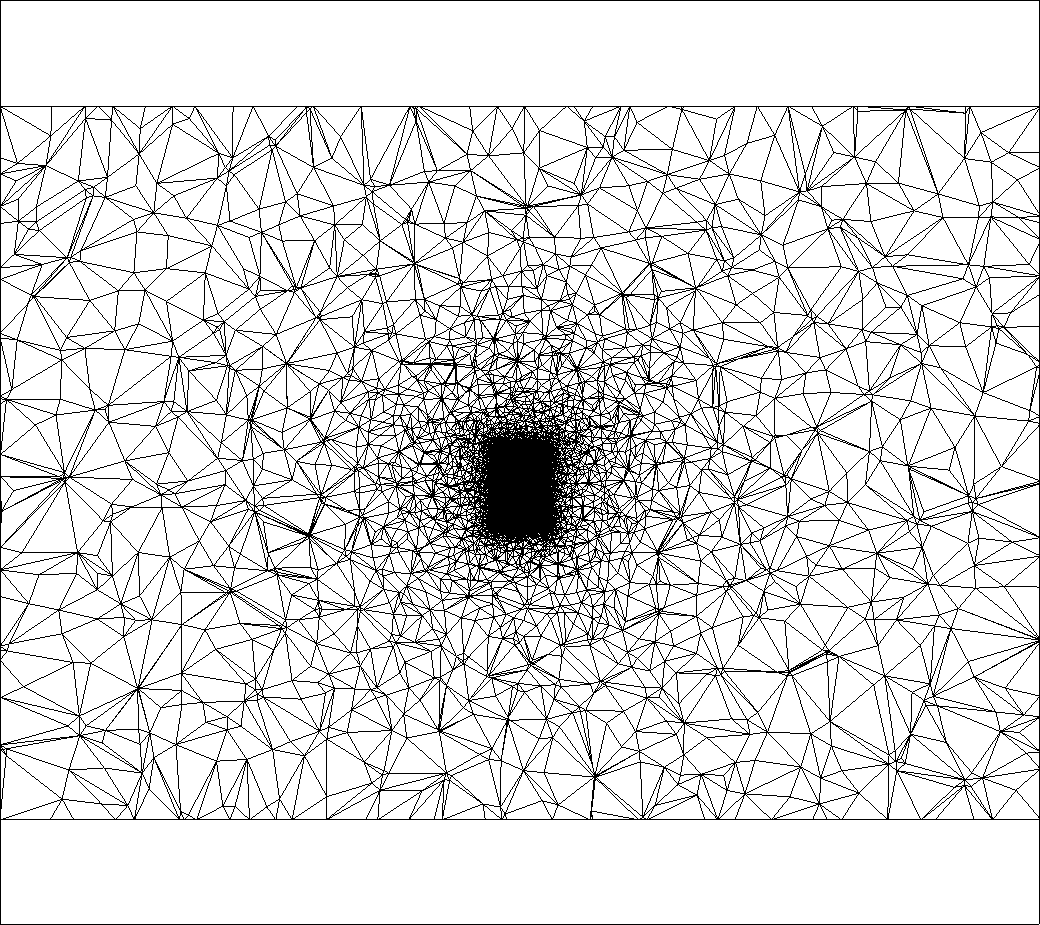}  
  \label{fig:entireMesh}
\end{subfigure}
\begin{subfigure}{.5\textwidth}
  \centering
  \includegraphics[width=.8\linewidth]{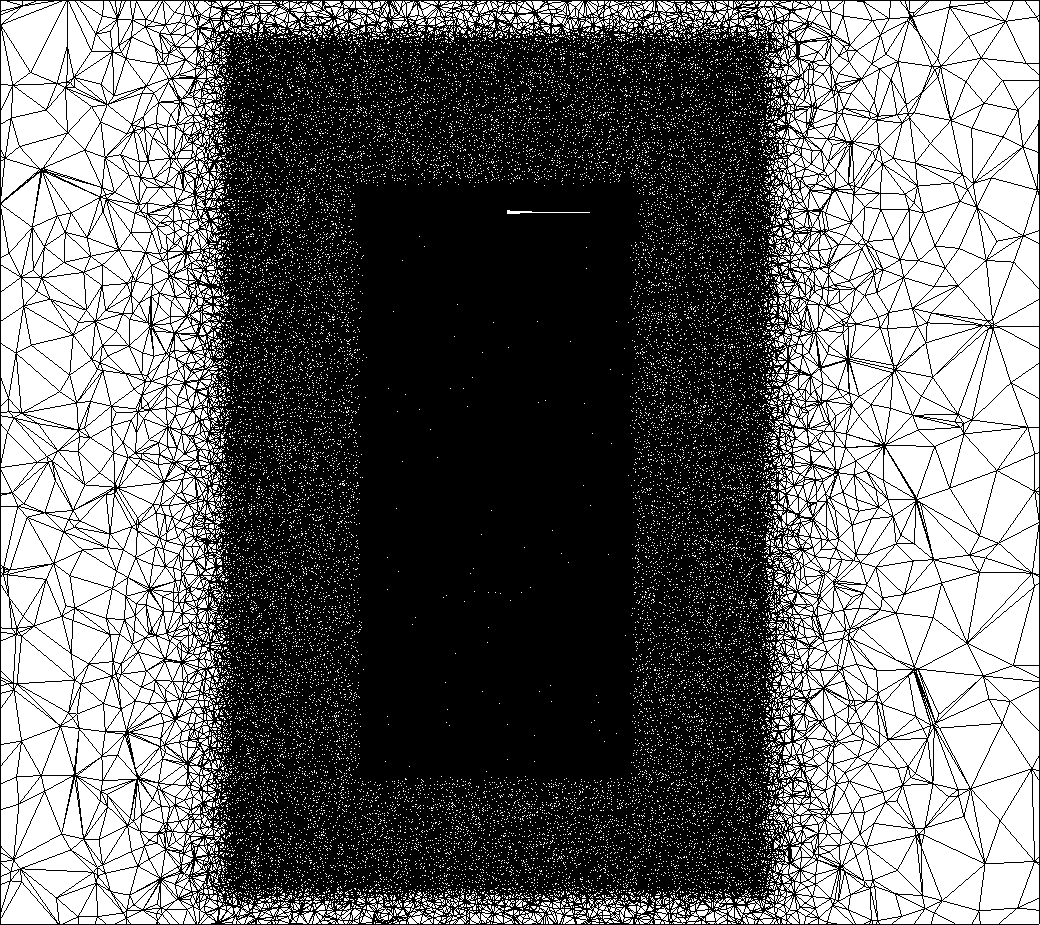}  
  \label{fig:middleZoomMesh}
\end{subfigure}
\newline
\begin{subfigure}{.5\textwidth}
  \centering
  \includegraphics[width=.8\linewidth]{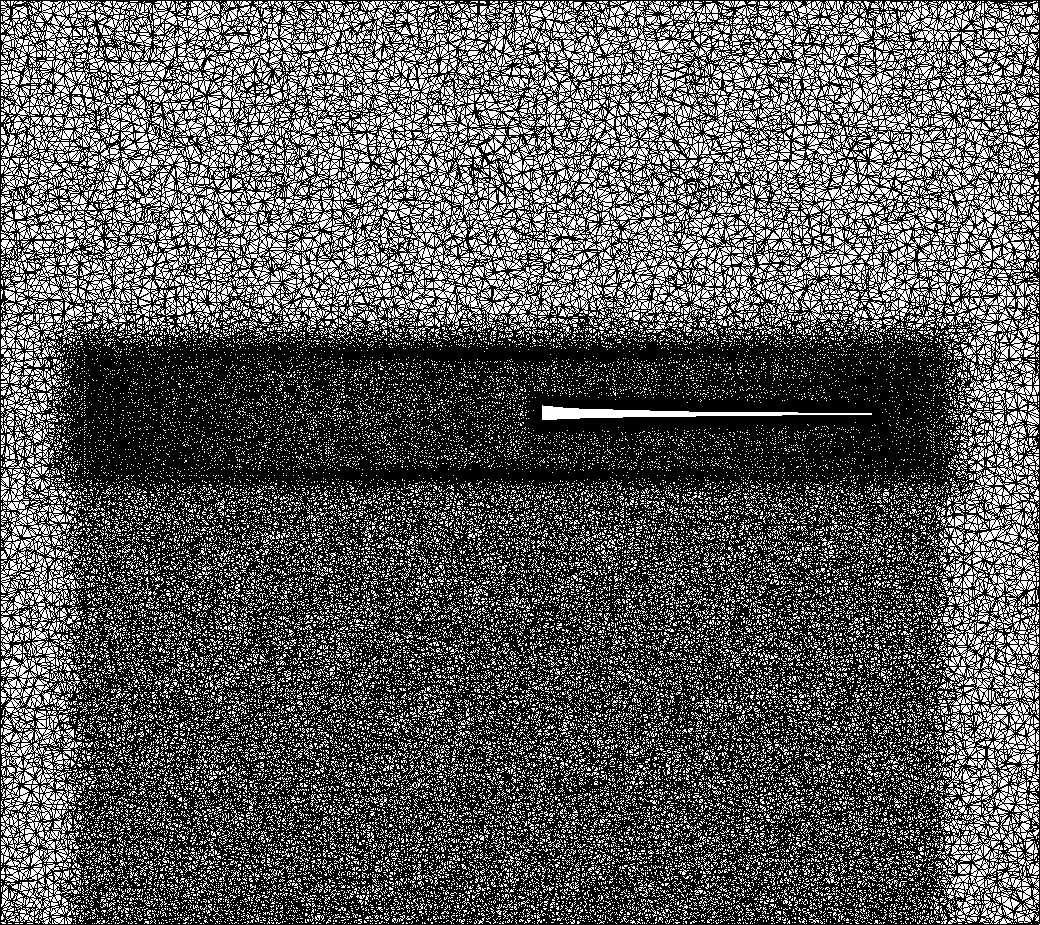}  
  \label{fig:smallZoomMesh}
\end{subfigure}
\begin{subfigure}{.5\textwidth}
  \centering
  \includegraphics[width=.8\linewidth]{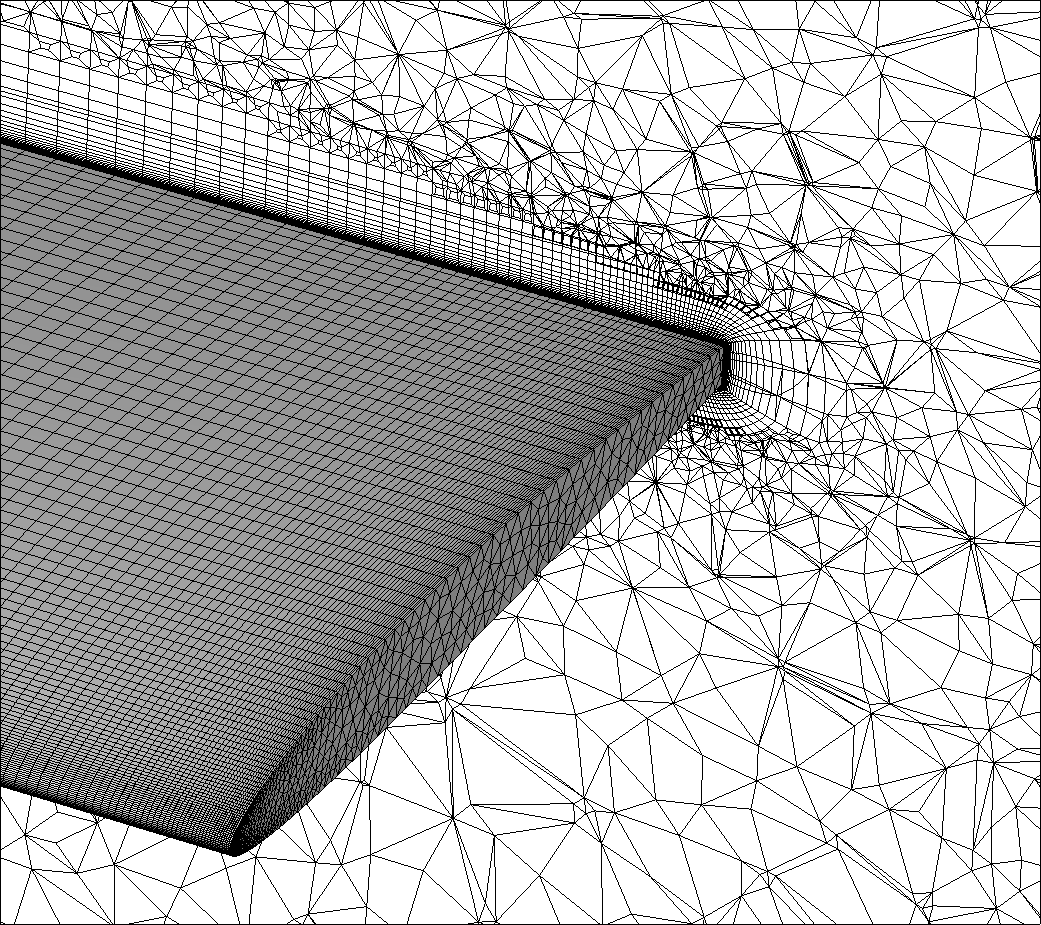}  
  \label{fig:meshtip}
\end{subfigure}
\caption{Cutting planes of volumetric mesh of 3-bladed XV-15 computational domain and surface mesh near blade tip.}
\label{fig:meshesSketch}
\end{figure}

More than 40 layers of Hexahedral cells are used near the blade to resolve the viscous boundary layer. Beyond that, tetrahedron cells are used in most places, while prismatic and pyramidal cells are used in between. The surface mesh points are directionally clustered near the blade leading edge, trailing edge and tip regions. The number of nodes on each blade’s surface is 58,641. The height of the first mesh layer above the blade surface was set to \num{7.143e-7} $\text{Chord}_{ref}$, which yields the $y^+$ less than 1.0 all over the blade. Figure~\ref{fig:meshesSketch} shows the volumetric grid distributions and the surface mesh resolutions as well as the boundary-layer mesh near the blade tip. The number of nodes in the farfield/nearfield background meshes and the body-fitted mesh are listed in Table~\ref{tab:meshing}

\begin{table}[H]
\centering
\caption{\label{tab:meshing}Meshing parameters for the XV-15 rotor mesh}
\begin{tabularx}{0.8\textwidth} { 
  | >{\raggedright\arraybackslash}X 
  | >{\centering\arraybackslash}X | }
 \hline
  & \textbf{Number of Nodes} \\
 \hline
 Blade mesh &  6.27M  \\
\hline
 Nearfield background mesh  & 2.69M    \\
\hline
 Farfield background mesh   &  15.10M  \\
\hline
\end{tabularx}
\end{table}

\section{The Numerical Methods}
\label{section:numericalMethods}
\begin{figure}[H]
    \centering
    \includegraphics[width=0.5\textwidth]{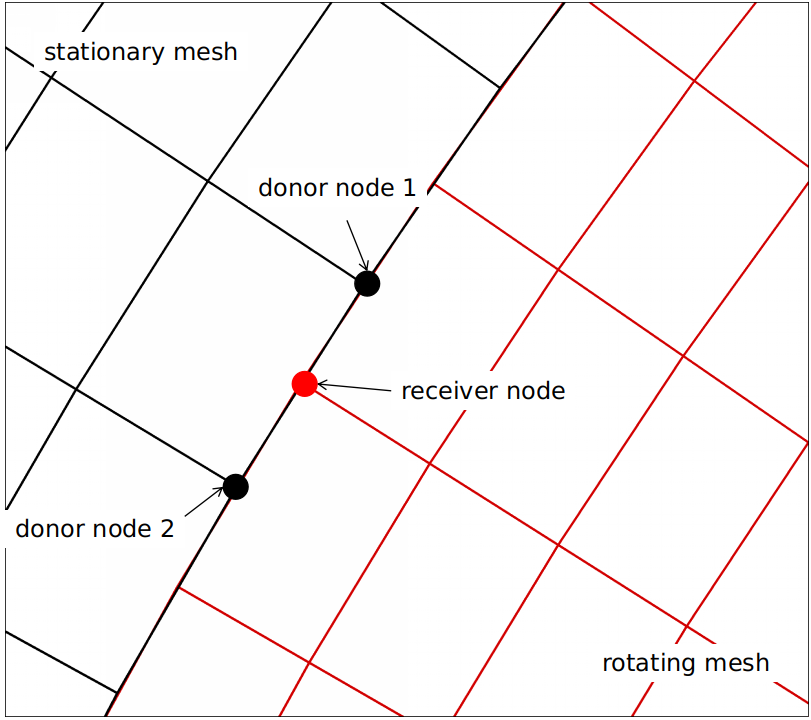}
    \caption{Overview of data communication from stationary mesh to rotating mesh on a sliding interface.}
    \label{fig:slidingInterface}
\end{figure}

We used Detached Eddy Simulation (DES) with Spalart–Allmaras (SA) turbulence model in the present study \cite{Spalart2009Detached-eddySimulation}. Figure \ref{fig:slidingInterface} shows the approach in which the solution is communicated from the stationary mesh to the rotating mesh on the sliding interface. At any moment during the rotation, every single receiver node on rotating mesh detects its two 'neighboring' nodes as donor nodes. The receiver node's solution is linearly interpolated from solutions of the two donor nodes. The communication from rotating mesh to stationary mesh is in the same manner.

\section{Test Conditions}
\label{section:testConditions}
Table~\ref{tab:flowCondition} summarises the flow conditions employed in the present study.

\begin{table}[H]
\centering
\caption{\label{tab:flowCondition}Flow conditions for the full-scale XV-15 tiltrotor blade.}
\begin{tabularx}{1.0\textwidth} { 
  | >{\raggedright\arraybackslash}X 
  | >{\centering\arraybackslash}X 
  | >{\centering\arraybackslash}X 
  | >{\raggedleft\arraybackslash}X |}
 \hline
 & \textbf{Hover Mode} & \textbf{Airplane Mode} &\textbf{Forward Mode} \\
 \hline
 Blade-tip Mach number ($M_{tip}$) & 0.69  & 0.54 & 0.69 \\
\hline
 Reynolds number ($Re$) & \num{4.95e6}  & \num{4.50e6} & \num{5.65e6} \\
\hline
 Blade pitch angle ($\theta_{75}$)& 0\degree,3\degree,5\degree,10\degree,13\degree & 26\degree,27\degree,28\degree,28.8\degree & 2\degree$\sim$10\degree \\
\hline
 Angle of attack ($\alpha$)& \backslashbox[0pt][lr]{}{} & -90\degree & -5\degree,0\degree,5\degree \\
\hline
 Advance ratio & \backslashbox[0pt][lr]{}{} & 0.337 & 0.170\\
\hline
 Grid size (\# of nodes) &  36.6M   & 36.6M   & 39.83M \\
\hline
\end{tabularx}
\end{table}

\section{Results and Discussions}
\label{section:results}
\subsection{Hovering Flight of Helicopter Mode}
\subsubsection{Overall Blade Loads}
\begin{figure}[H]
    \centering
    \includegraphics[width=0.75\textwidth]{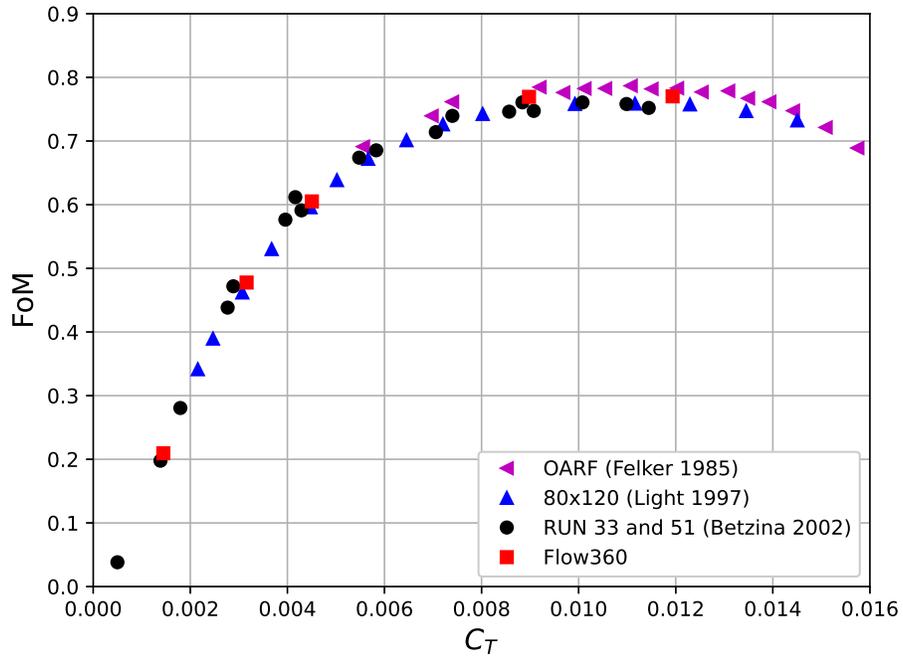}
    \caption{Figure of merit - Thurst coefficient of hovering helicopter mode.}
    \label{fig:ct_fom_hover}
\end{figure}

\begin{figure}[H]
    \centering
    \includegraphics[width=0.75\textwidth]{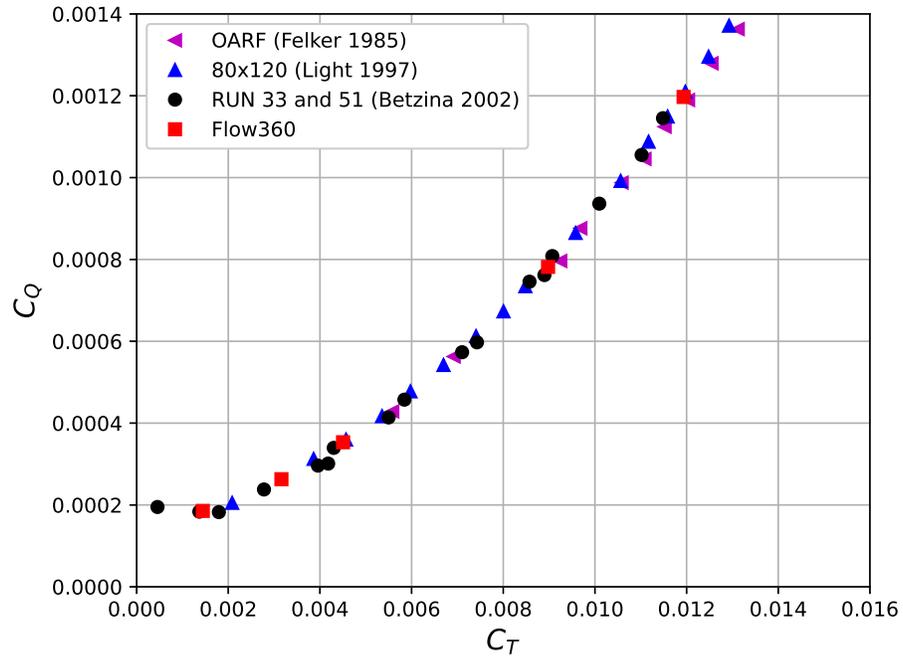}
    \caption{Torque coefficient - Thrust coefficient of hovering helicopter mode}
    \label{fig:ct_cq_hover}
\end{figure}

The figure of merit and torque coefficient $C_Q$, as functions of the thrust coefficient $C_T$, are shown in Figure~\ref{fig:ct_fom_hover} and Figure~\ref{fig:ct_cq_hover} respectively. Three sets of experimental data of the full-scale XV-15 rotor are also shown, carried out by Felker et al.~\cite{Felker1986PerformanceRotor} at OARF, and Light~\cite{Light1997ResultsComplex} and Betzina~\cite{Betzina2002RotorMode} at the NASA 80$\times$120ft wind tunnel. The results from Flow360 present a good agreement with the experimental data. 

\subsubsection{Surface Pressure Coefficient and Skin Friction}
\begin{figure}[H]
\begin{subfigure}{.5\textwidth}
  \centering
  \includegraphics[width=.8\linewidth]{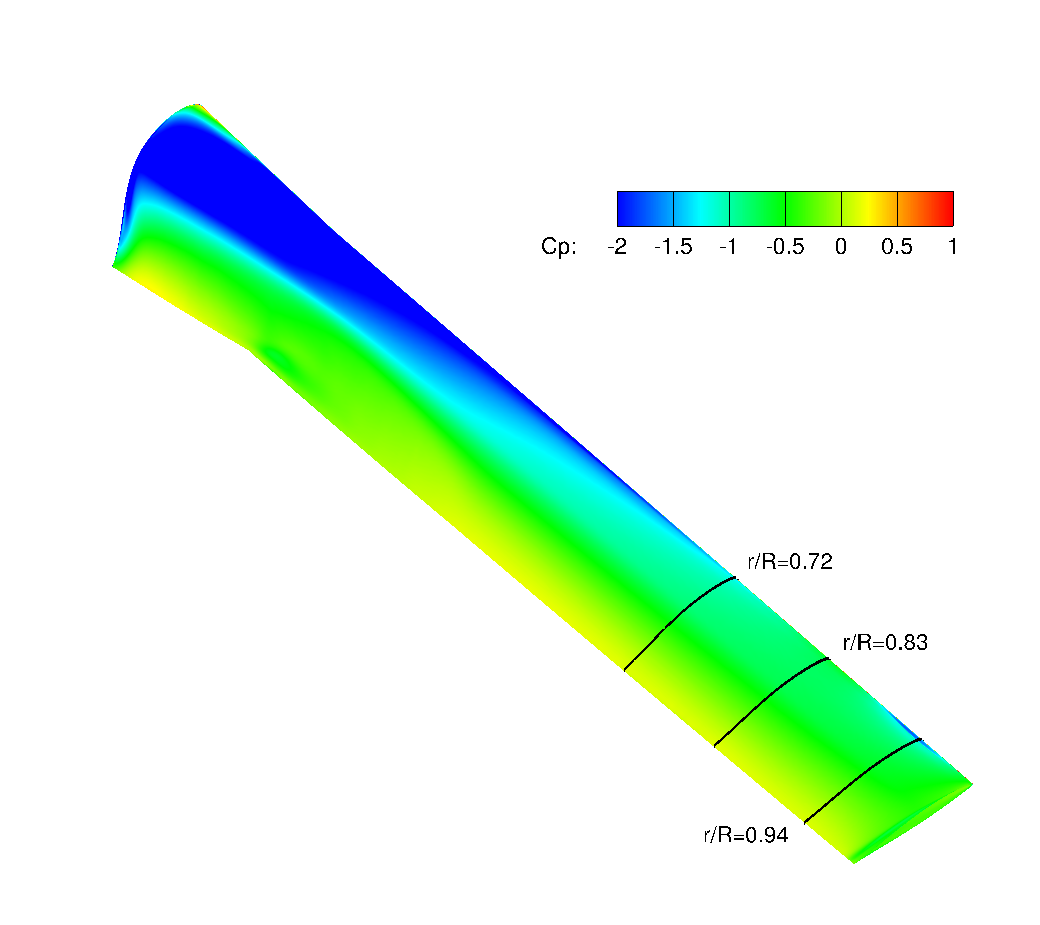}  
  \caption{Surface pressure coefficient}
  \label{fig:Cp_contour_pitch10}
\end{subfigure}
\begin{subfigure}{.5\textwidth}
  \centering
  \includegraphics[width=.8\linewidth]{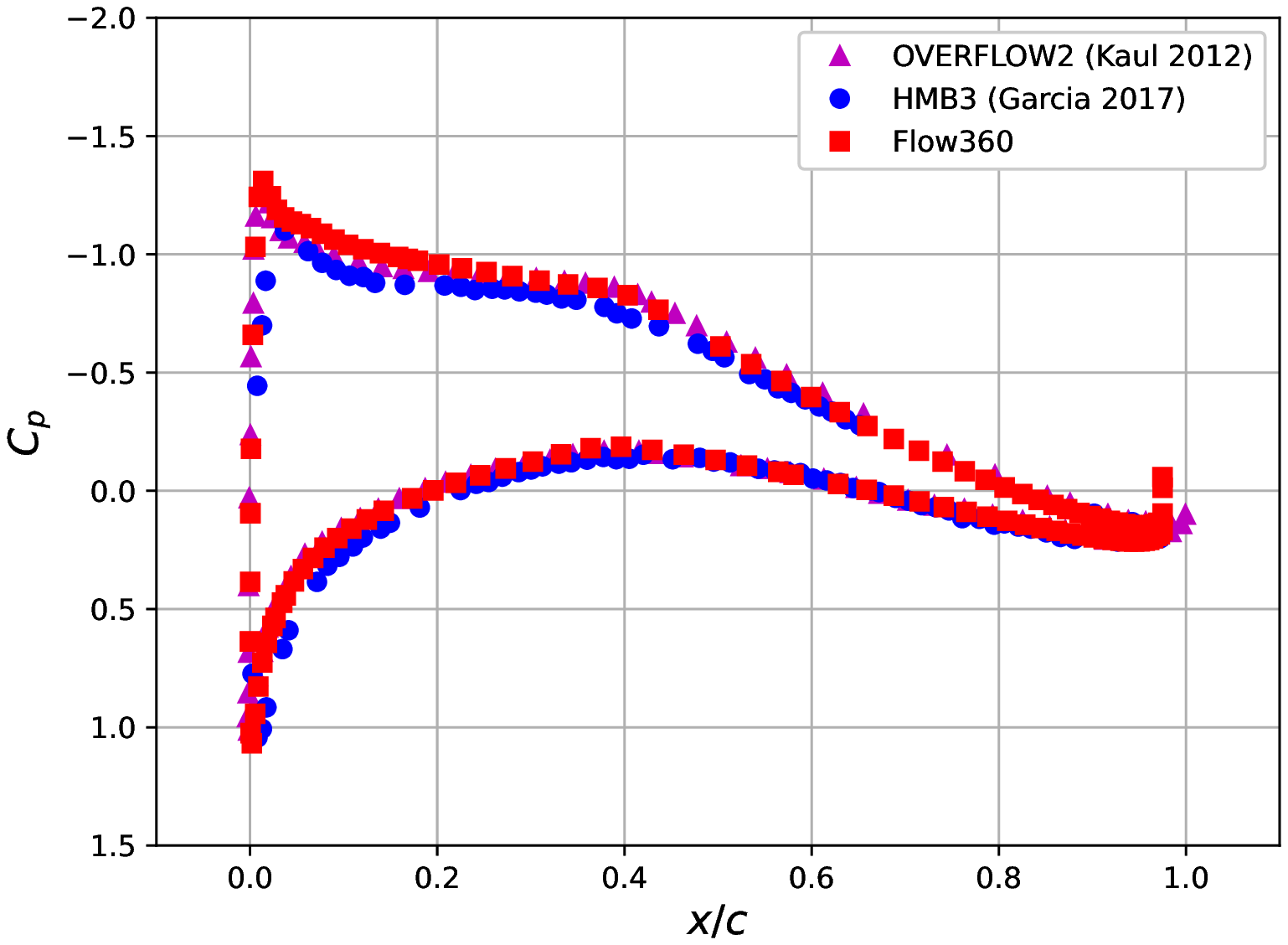}  
  \caption{$r/R$=0.72}
  \label{fig:Cp_0.72}
\end{subfigure}
\newline
\begin{subfigure}{.5\textwidth}
  \centering
  \includegraphics[width=.8\linewidth]{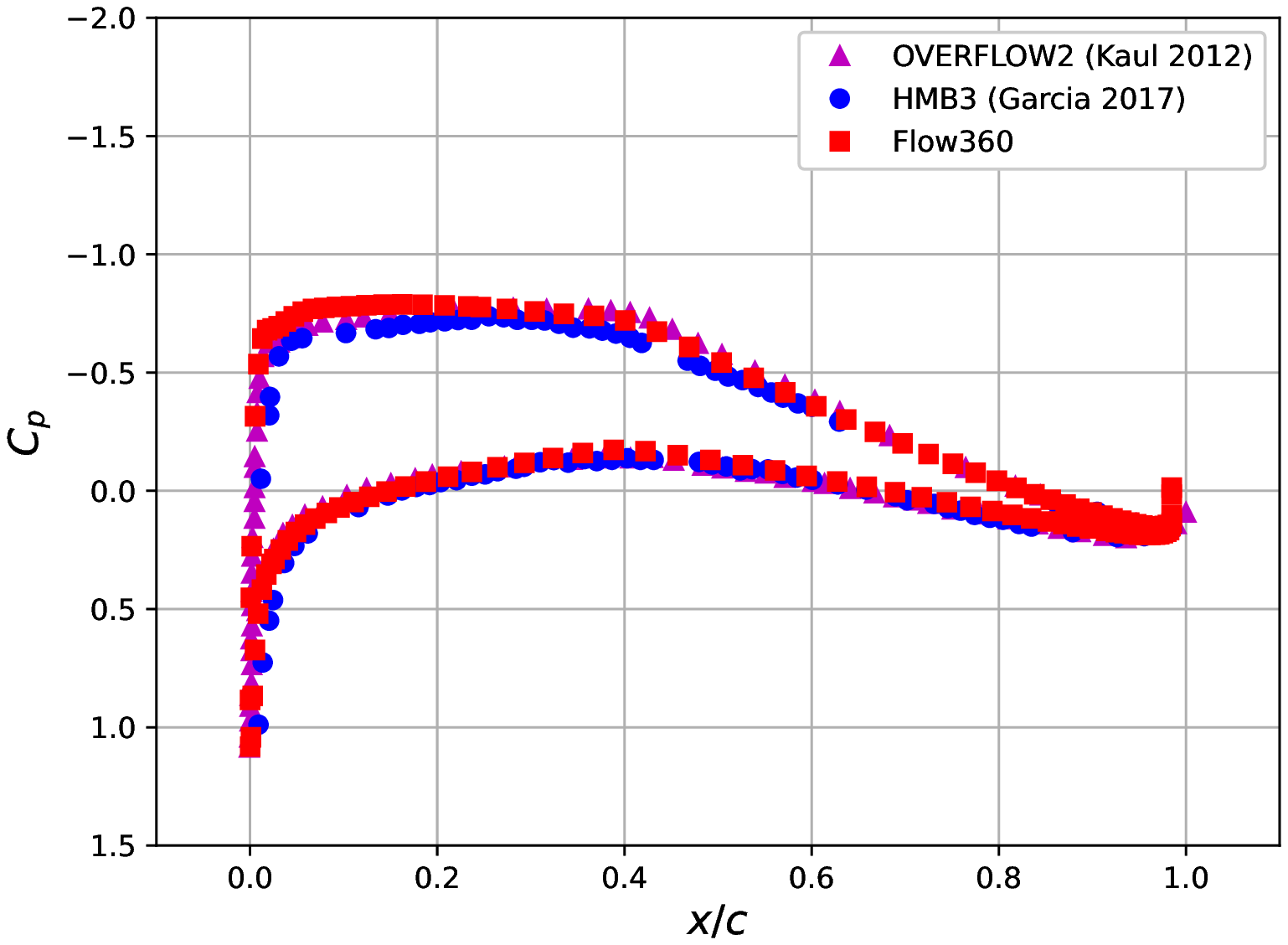}  
  \caption{$r/R$=0.83}
  \label{fig:Cp_0.83}
\end{subfigure}
\begin{subfigure}{.5\textwidth}
  \centering
  \includegraphics[width=.8\linewidth]{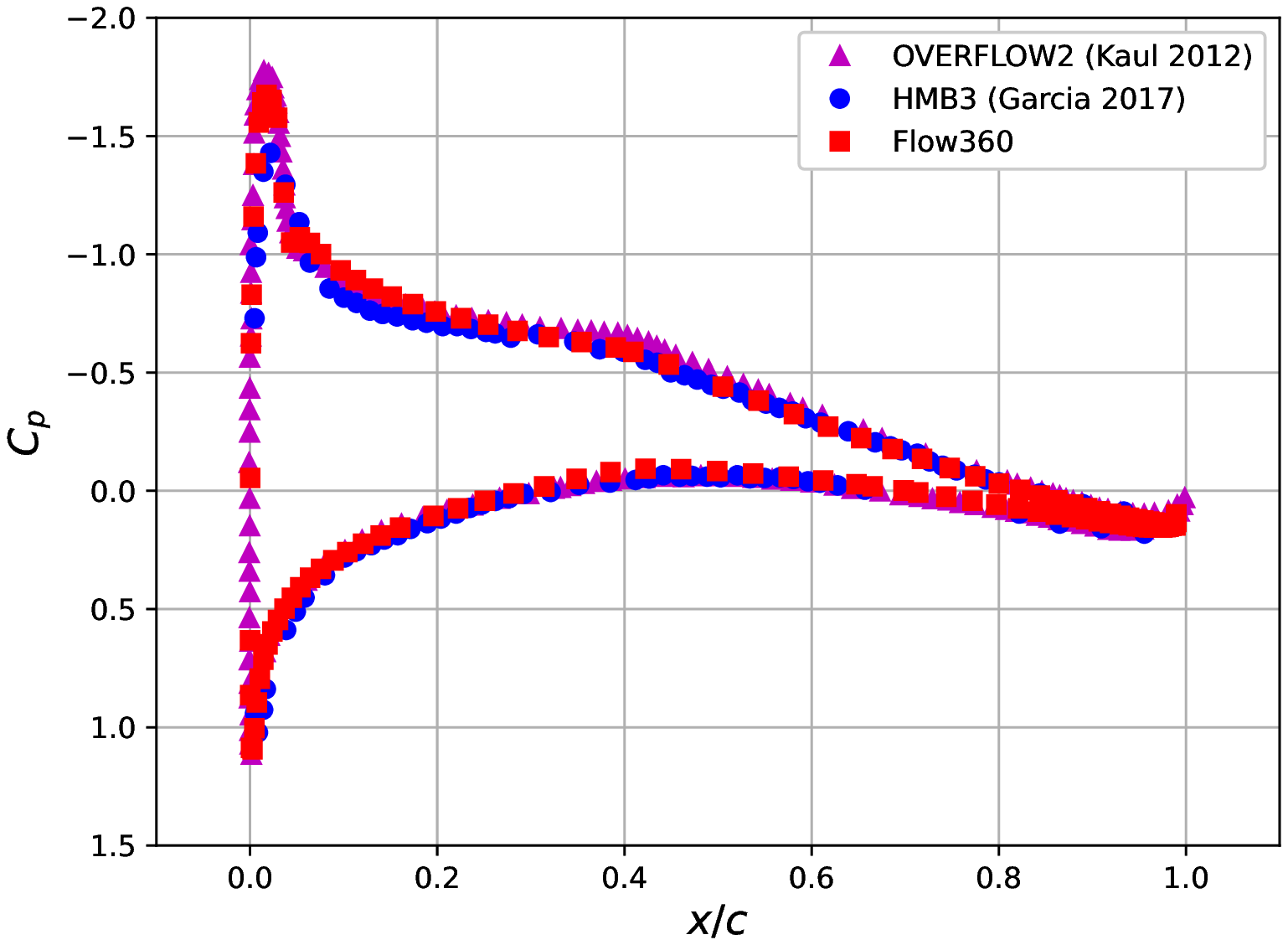}  
  \caption{$r/R$=0.94}
  \label{fig:Cp_0.94}
\end{subfigure}
\caption{Comparison of predicted surface pressure coefficient among Flow360, OVERFLOW2 and HMB3 at collective pitch angle $\theta_{75}=10\degree$.}
\label{fig:cp_comparison_pitch10}
\end{figure}

To validate the predictive capability of Flow360 in greater physical detail, surface pressure and viscous friction distributions are investigated. For surface pressure, because there is no experimental measurements on it, the CFD data by OVERFLOW2 \cite{Kaul2011SkinOVERFLOW2} and HMB3 \cite{Jimenez-Garcia2017TiltrotorValidation} is used as reference for comparison shown in Figure \ref{fig:cp_comparison_pitch10}. Three radial stations were selected for the collective pitch angle = 10\degree. The surface pressure coefficient is calculated based on the local rotating velocity at each radial station in Eq.\ref{definition_of_Cp}:
\begin{equation}
\label{definition_of_Cp}
C_P=\frac{P-P_\infty}{1/2\rho_\infty\bigl(\Omega r\bigr)^2}
\end{equation}

\begin{figure}[H]
\begin{subfigure}{.5\textwidth}
  \centering
  \includegraphics[width=.8\linewidth]{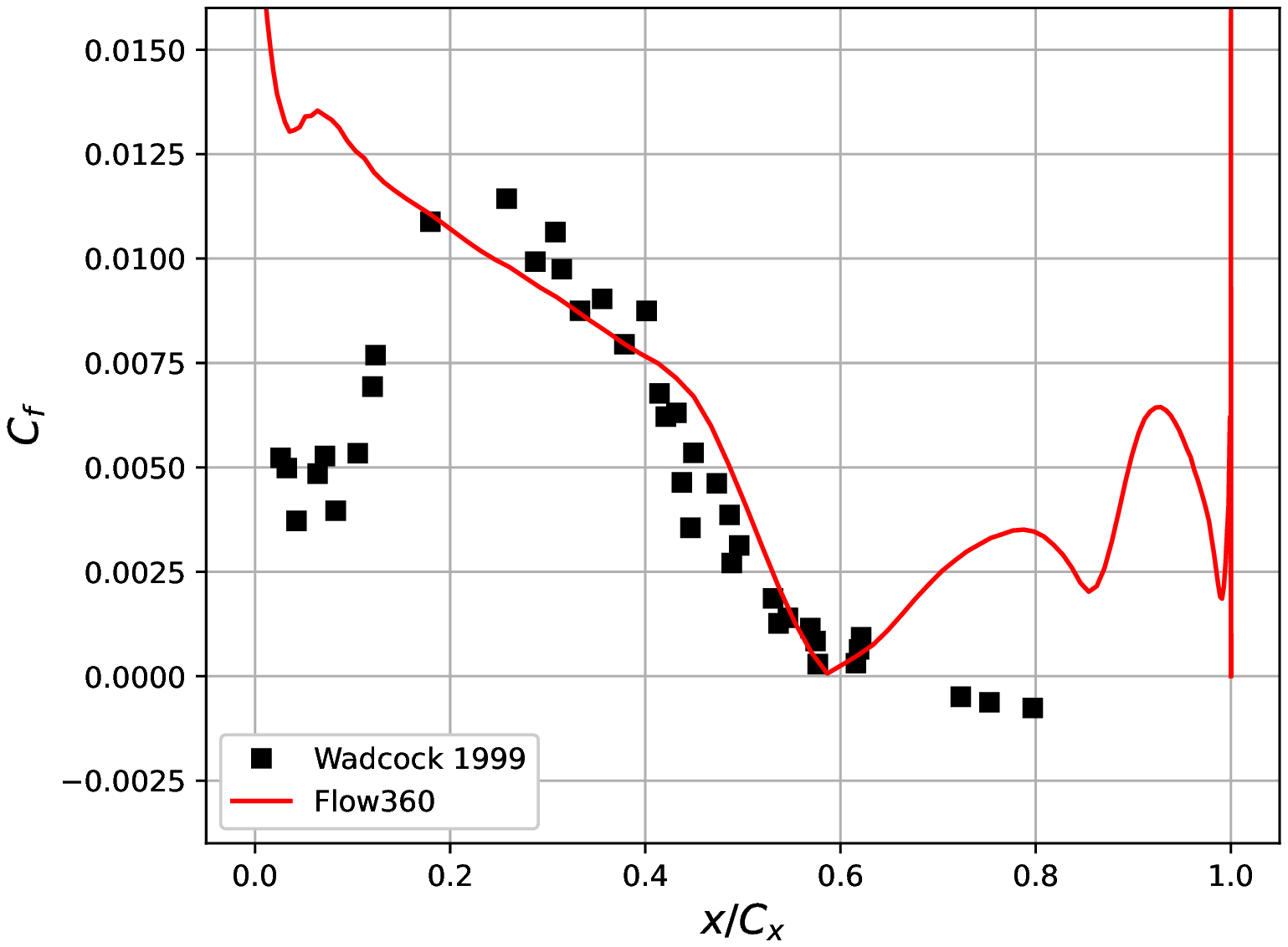}  
  \caption{$r/R=0.28$}
  \label{fig:Cf_pitch10_0.28}
\end{subfigure}
\begin{subfigure}{.5\textwidth}
  \centering
  \includegraphics[width=.8\linewidth]{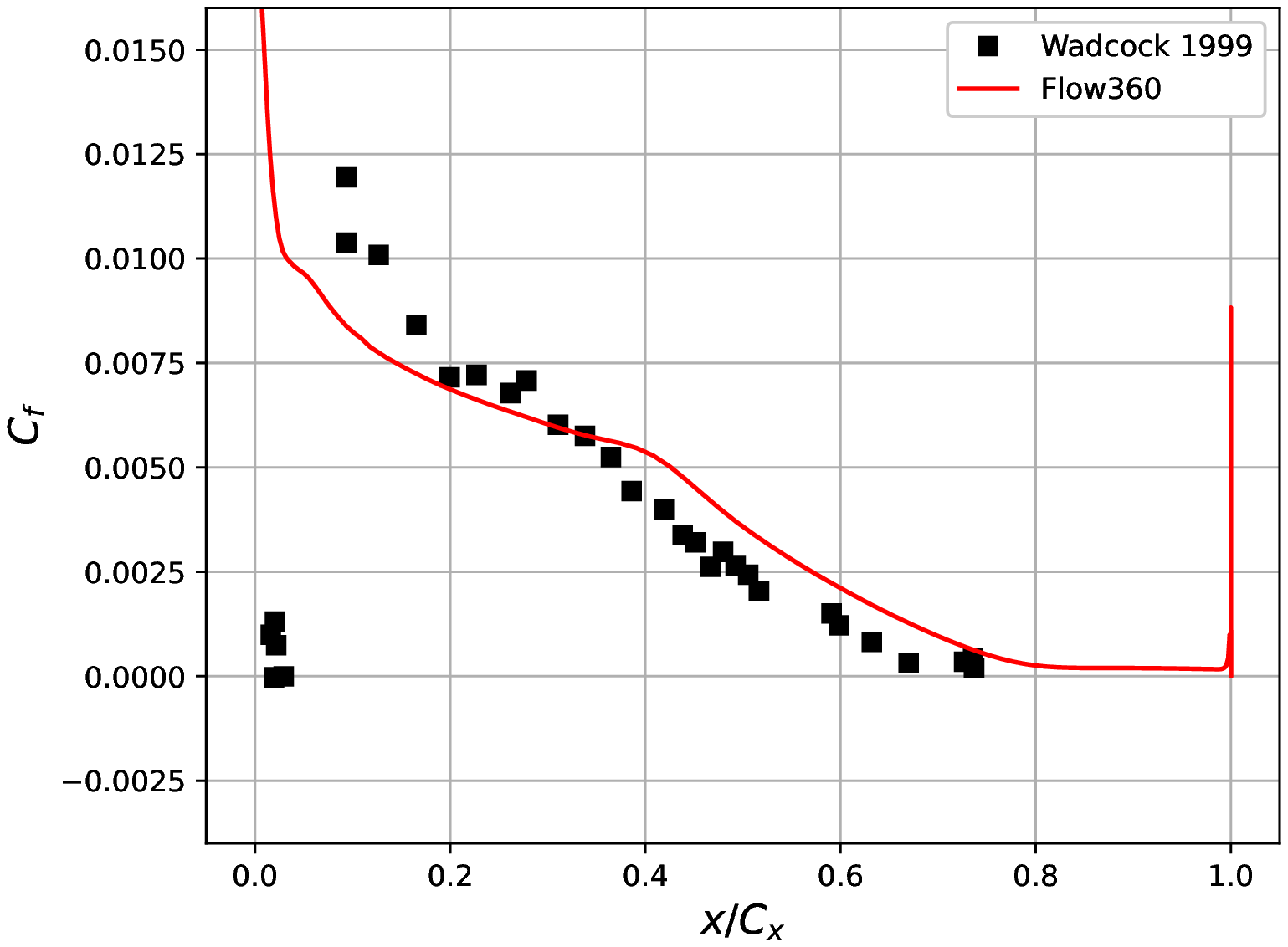}  
  \caption{$r/R=0.50$}
  \label{fig:Cf_pitch10_0.50}
\end{subfigure}
\newline
\begin{subfigure}{.5\textwidth}
  \centering
  \includegraphics[width=.8\linewidth]{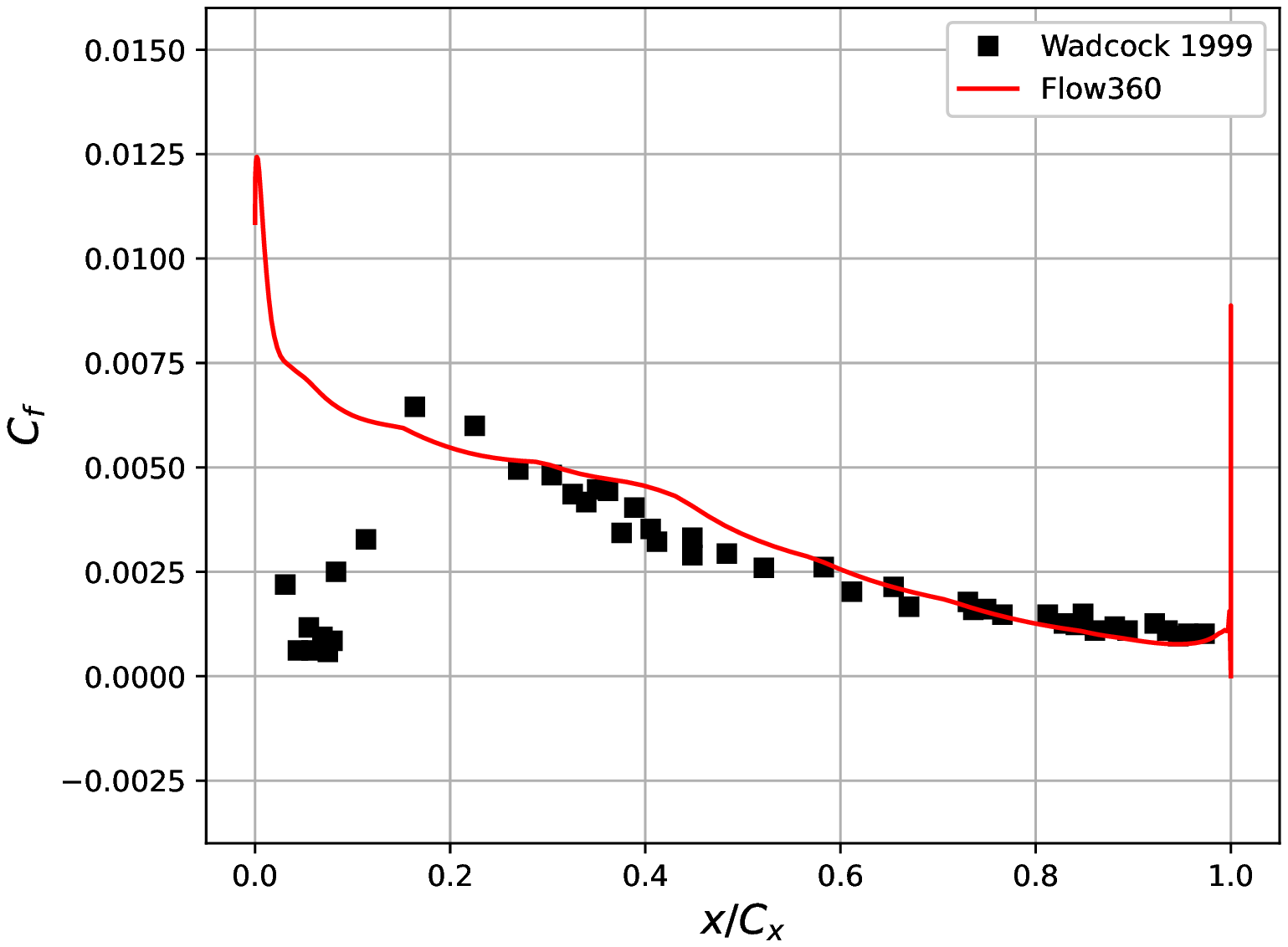}  
  \caption{$r/R=0.72$}
  \label{fig:Cf_pitch10_0.72}
\end{subfigure}
\begin{subfigure}{.5\textwidth}
  \centering
  \includegraphics[width=.8\linewidth]{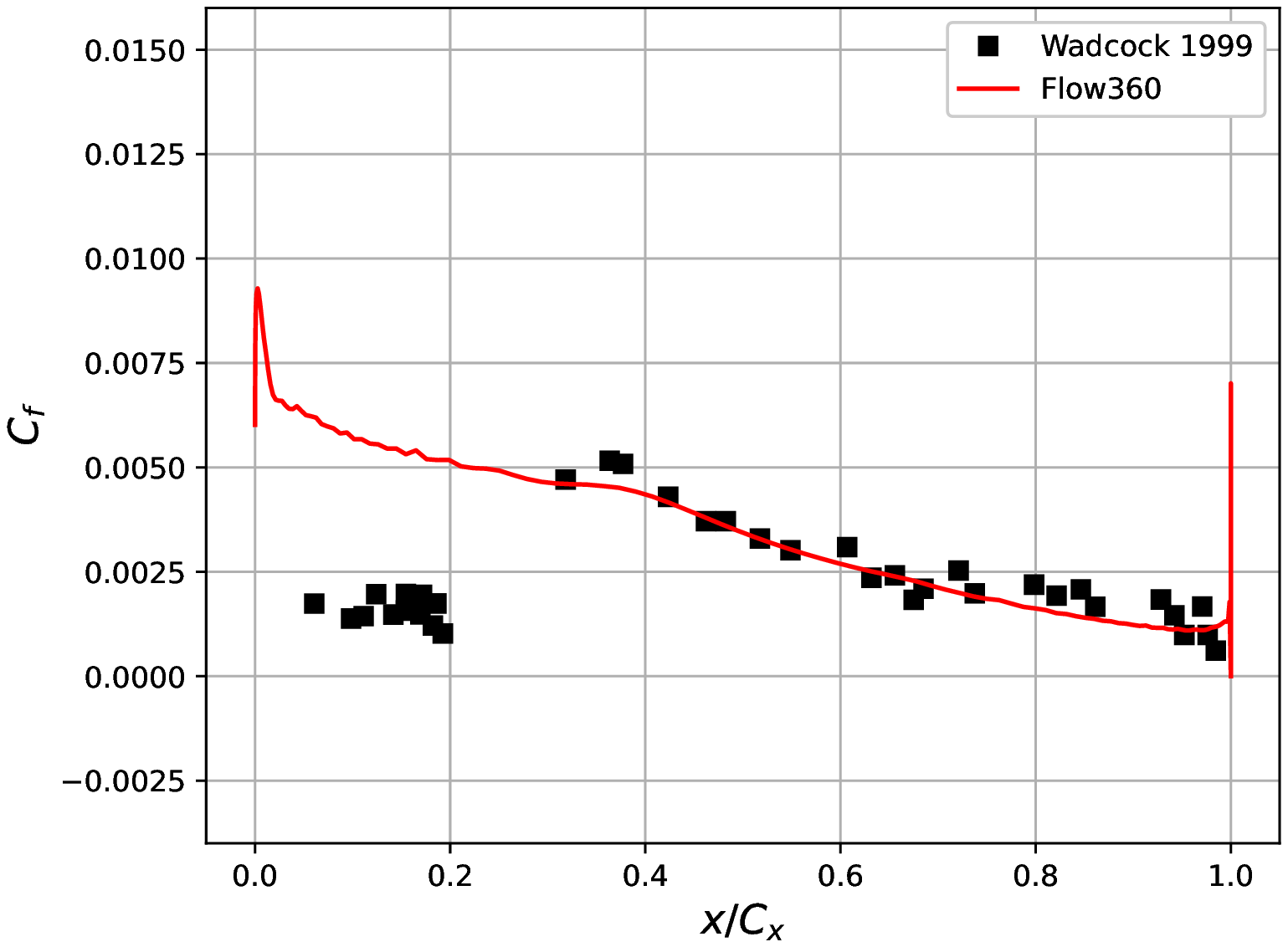}  
  \caption{$r/R=0.83$}
  \label{fig:Cf_pitch10_0.83}
\end{subfigure}
\newline
\begin{subfigure}{.5\textwidth}
  \centering
  \includegraphics[width=.8\linewidth]{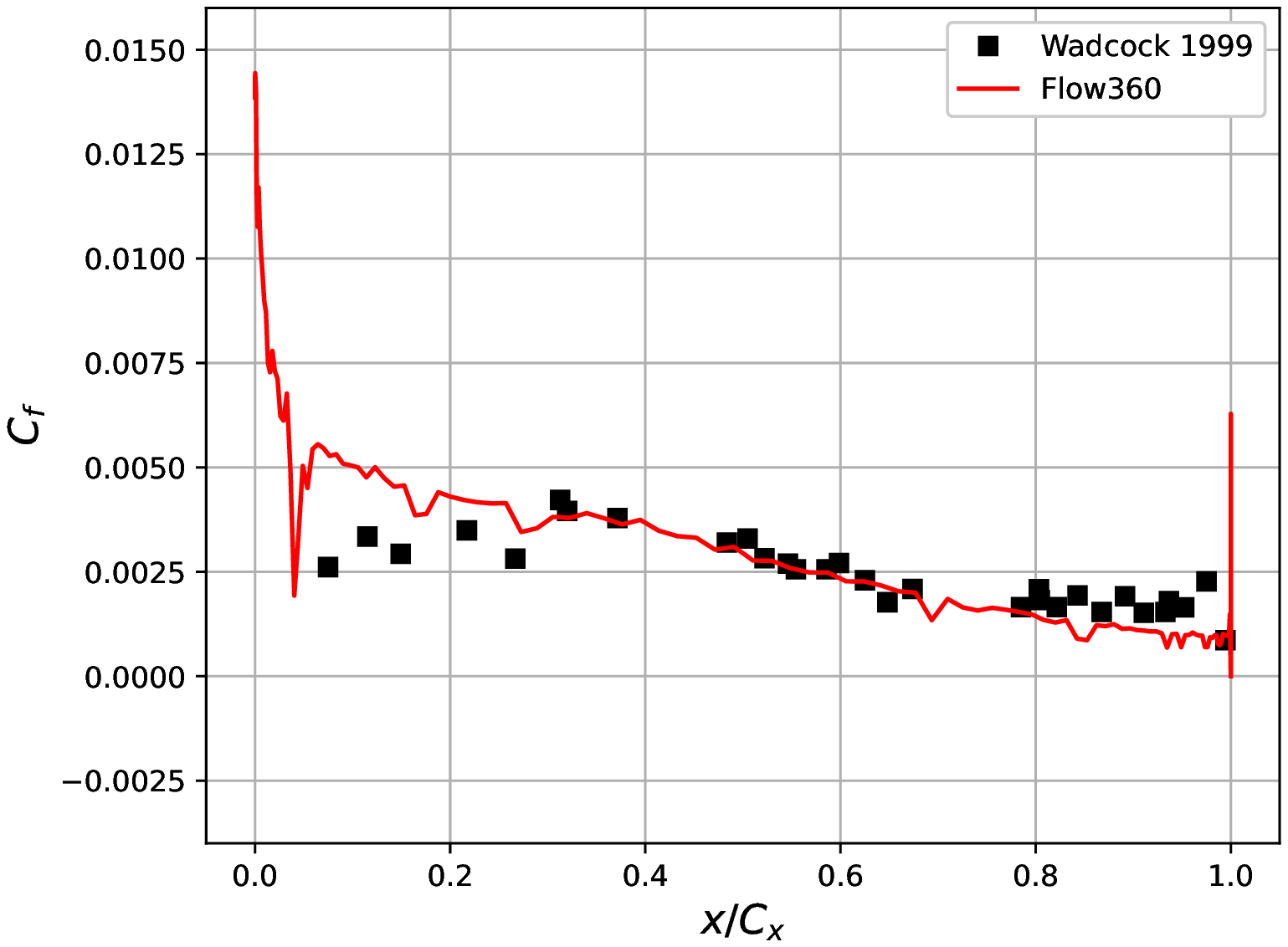}  
  \caption{$r/R=0.94$}
  \label{fig:Cf_pitch10_0.94}
\end{subfigure}
\caption{Comparison of predicted skin friction coefficient with the experimental data of Wadcock et al.\cite{Wadcock1999SkinRotor}. Conditions employed: $M_{tip}=0.69$, $Re=4.95\cdot 10^6$, and $\theta_{75}=10\degree$.}
\label{fig:cf_comparison_pitch10}
\end{figure}

In order to provide finer assessment of the Flow360 in capturing physics phenomenon, the computed skin friction coefficients, defined in Eq.\ref{eq:definition_of_Cf}, with experimental data by Wadcock et al.~\cite{Wadcock1999SkinRotor} at 6 blade span stations: $r/R$ = 0.17, 0.28, 0.50, 0.72, 0.84 and 0.94 for collective pitch angle $\theta_{75}=10\degree$ are shown in Figure~\ref{fig:cf_comparison_pitch10}. 
\begin{equation}
\label{eq:definition_of_Cf}
C_f=\frac{\tau_{wall}}{1/2\rho_\infty\bigl(\Omega r\bigr)^2}
\end{equation}
At the outboard station $r/R$ = 0.94, the flow is fully turbulent shown by the measured data and the computed $C_f$ matches the measured data well. At other stations, the experimental data shows natural transition phenomenon. In turbulent flow regions of these stations, the computed $C_f$ matches the measured data well, except the region $x/C_x>0.6$ at the station $r/R$=0.28, which will be investigated further. In laminar regions, because no transition model was employed in the present study, the computed $C_f$ over predicts the skin friction as expected. This result shows the capability of capturing the natural transition phenomenon is crucial to accurately predict the skin friction on the rotor blades, which is one of the future works.

\subsubsection{Sectional Blade Loads}
\begin{figure}[H]
\begin{subfigure}{.5\textwidth}
  \centering
  \includegraphics[width=.8\linewidth]{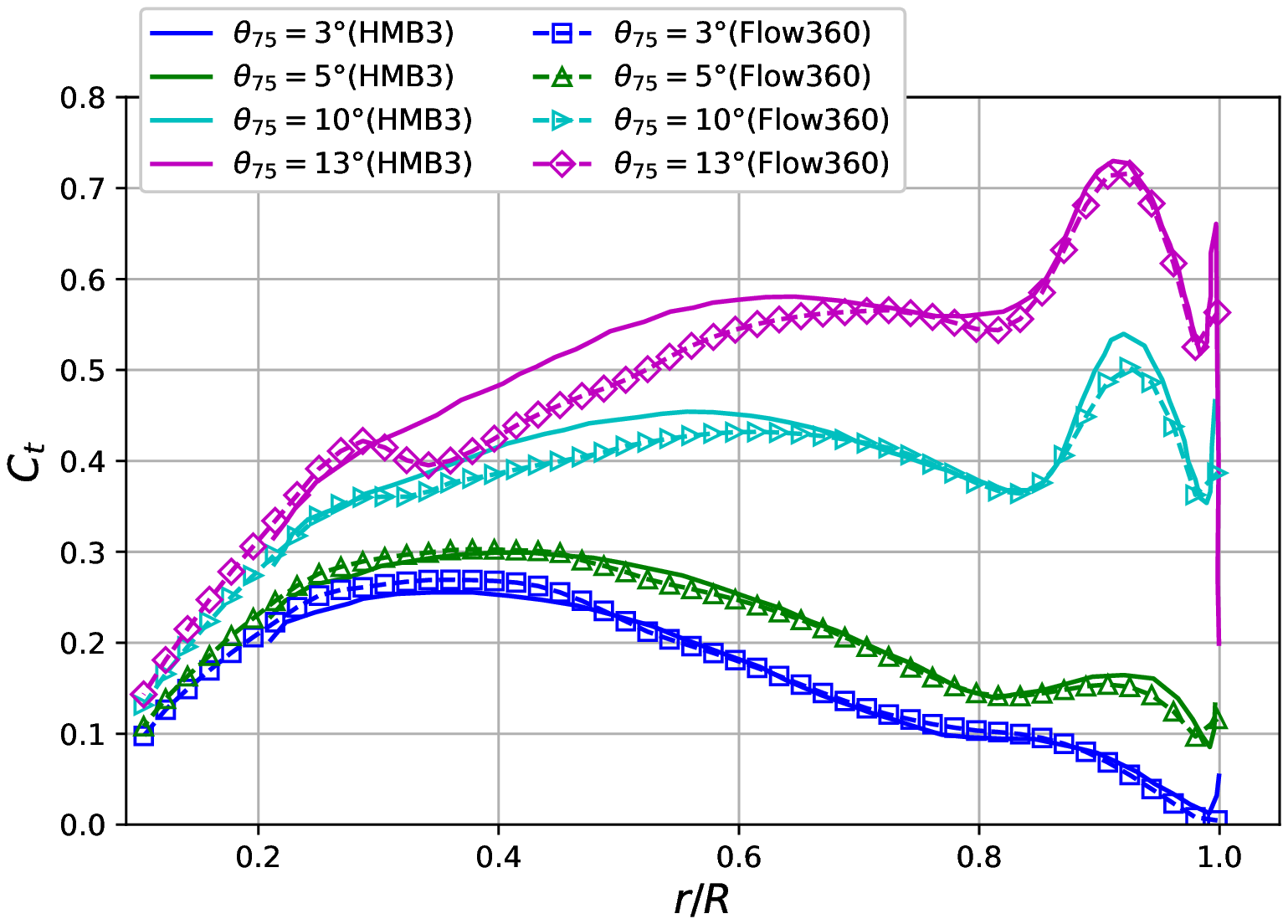}  
  \caption{Sectional thrust coefficient}
  \label{fig:sectional_ct_hover}
\end{subfigure}
\begin{subfigure}{.5\textwidth}
  \centering
  \includegraphics[width=.8\linewidth]{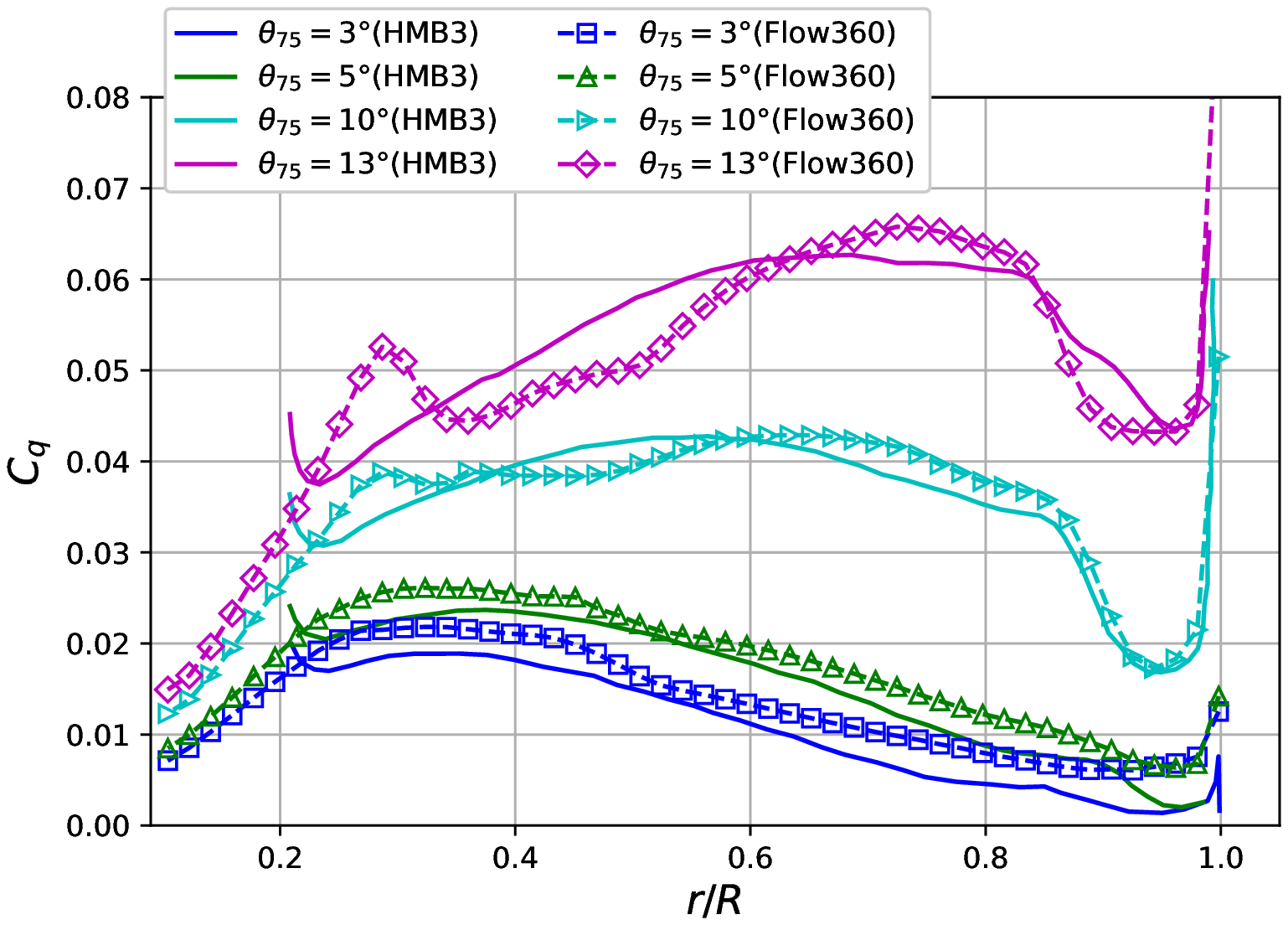}
  \caption{Sectional torque coefficient}
  \label{fig:sectional_cq_hover}
\end{subfigure}
\caption{Blade sectional thrust and torque coefficient for the full-scale XV-15 rotor in helicopter mode.}
\label{fig:sectional_ctcq_hover}
\end{figure}

To further understand the flow physics near the rotor blades, the sectional loading was investigated. The sectional thrust coefficient $C_t$ and torque coefficient $C_q$ are distributions along the radius $r$. They are defined in Eq.\ref{definition_of_sectional_Ct} and Eq.\ref{definition_of_sectional_Cq} respectively.
\begin{equation}
\label{definition_of_sectional_Ct}
C_t\bigl(r\bigr)=\frac{\int\bigl((p-p_\infty)\vec{n}+\bm{\tau}\cdot\vec{n}\bigr)_{axial}\,ds}{\frac{1}{2}\rho_{\infty}\left((\Omega r)^2+V_{\infty}^2\right)\text{chord}_{\text{local}}}\cdot\frac{r}{R}
\end{equation}
\begin{equation}
\label{definition_of_sectional_Cq}
C_q\bigl(r\bigr)=\frac{\int\left[\bigl((p-p_\infty)\vec{n}+\bm{\tau}\cdot\vec{n}\bigr)\times \vec{r}\right]_{axial}\,ds}{\frac{1}{2}\rho_{\infty}\left((\Omega r)^2+V_{\infty}^2\right)\text{chord}_{\text{local}}R}\cdot\frac{r}{R}
\end{equation}
Due to the lack of experimental data on the sectional loading, CFD data of HMB3 \cite{Jimenez-Garcia2017TiltrotorValidation} is shown as comparison for multiple collective pitch angles $\theta_{75}=3\degree, 5\degree, 10\degree, 13\degree$ in Figure~\ref{fig:sectional_ctcq_hover}. It shows the two solvers agree well with each other in the entire span for all collective pitch angles. Both of them predict the spike of thrust near the tip, $r/R=0.9$, due to the blade vortex interaction.

\subsubsection{Flowfield Details}
\begin{figure}[H]
\centering
\includegraphics[width=.8\linewidth]{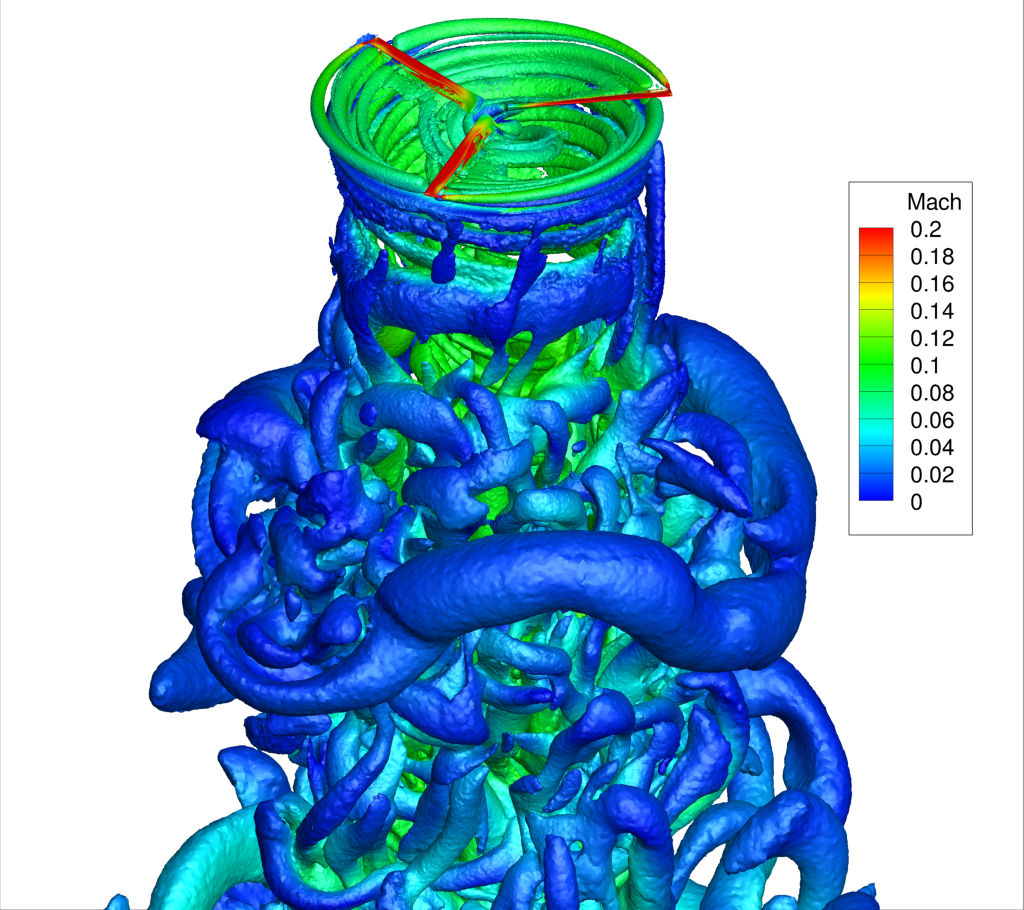}  
\caption{Wake visualization of the hovering XV-15 blades using Q-criterion shaded by contours of Mach at collective pitch $\theta_{75}=10\degree$.}
\label{fig:hover_qcriterion}
\end{figure}

Flowfield visualization of the rotor wake for the full-scale XV-15 rotor blade in hovering flight of helicopter mode using the Q criterion is given in Figure~\ref{fig:hover_qcriterion}. The interaction between blade and vortex can be seen evidently.

\subsection{Airplane Mode}
\subsubsection{Overall Blade Loads}
\begin{figure}[H]
    \centering
    \includegraphics[width=0.75\textwidth]{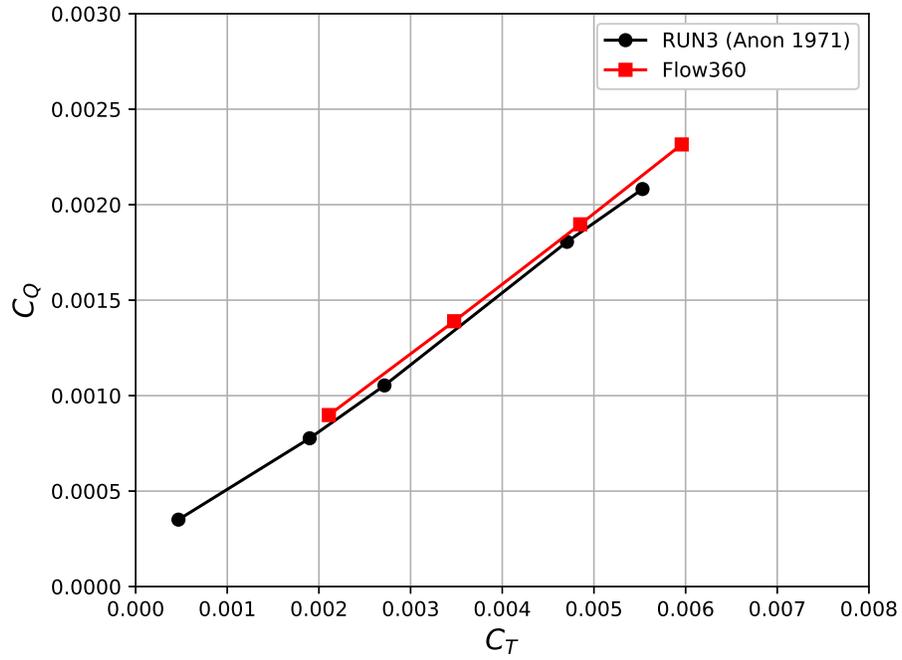}
    \caption{Torque coefficient - Thrust coefficient of airplane mode.}
    \label{fig:ct_cq_propeller}
\end{figure}

\begin{figure}[H]
    \centering
    \includegraphics[width=0.75\textwidth]{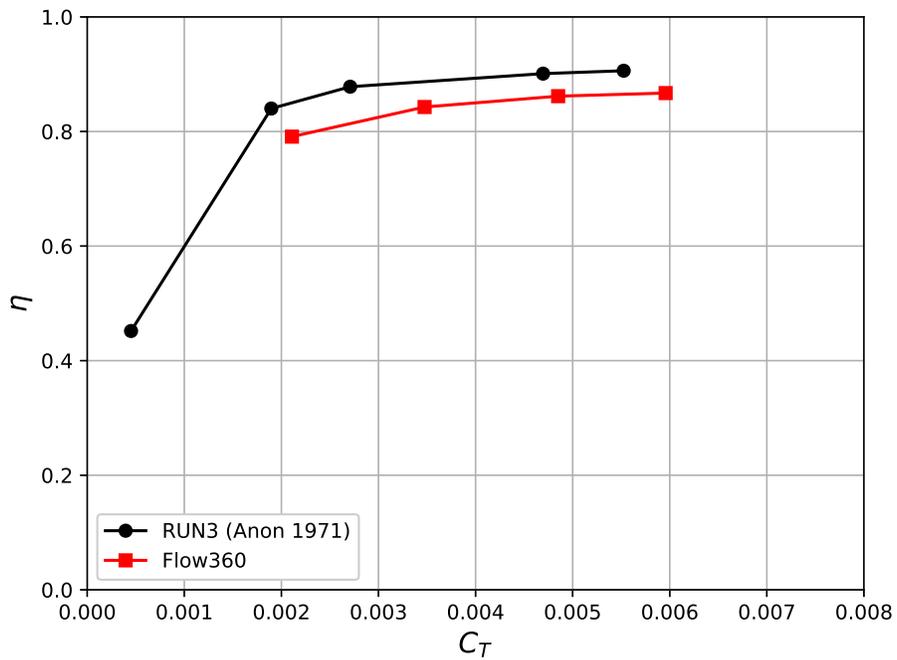}
    \caption{Propeller propulsive efficiency - Thrust coefficient of airplane mode.}
    \label{fig:ct_eta_propeller}
\end{figure}

For airplane mode, simulations were performed for medium advance ratio $\mu=0.337$ at collective pitch angles of 26\degree, 27\degree, 28\degree, 28.8\degree, and tip Mach number of 0.54. The $C_Q$ as a function of $C_T$ is shown in Fig.~\ref{fig:ct_cq_propeller}. In this mode, the indicator to measure the rotor efficiency is the propeller propulsive efficiency, defined as the ratio between the useful power output of the propeller and the absorbed power expressed in Eq.(\ref{definition_of_yita}). The $\eta$ as a function of $C_T$ is shown in Fig.\ref{fig:ct_eta_propeller}.
\begin{equation}
\label{definition_of_yita}
\eta=\frac{C_{T}V_{\infty}}{C_{Q}V_{tip}}
\end{equation}

\subsubsection{Surface Pressure Predictions}
\begin{figure}[H]
\begin{subfigure}{.5\textwidth}
  \centering
  \includegraphics[width=.8\linewidth]{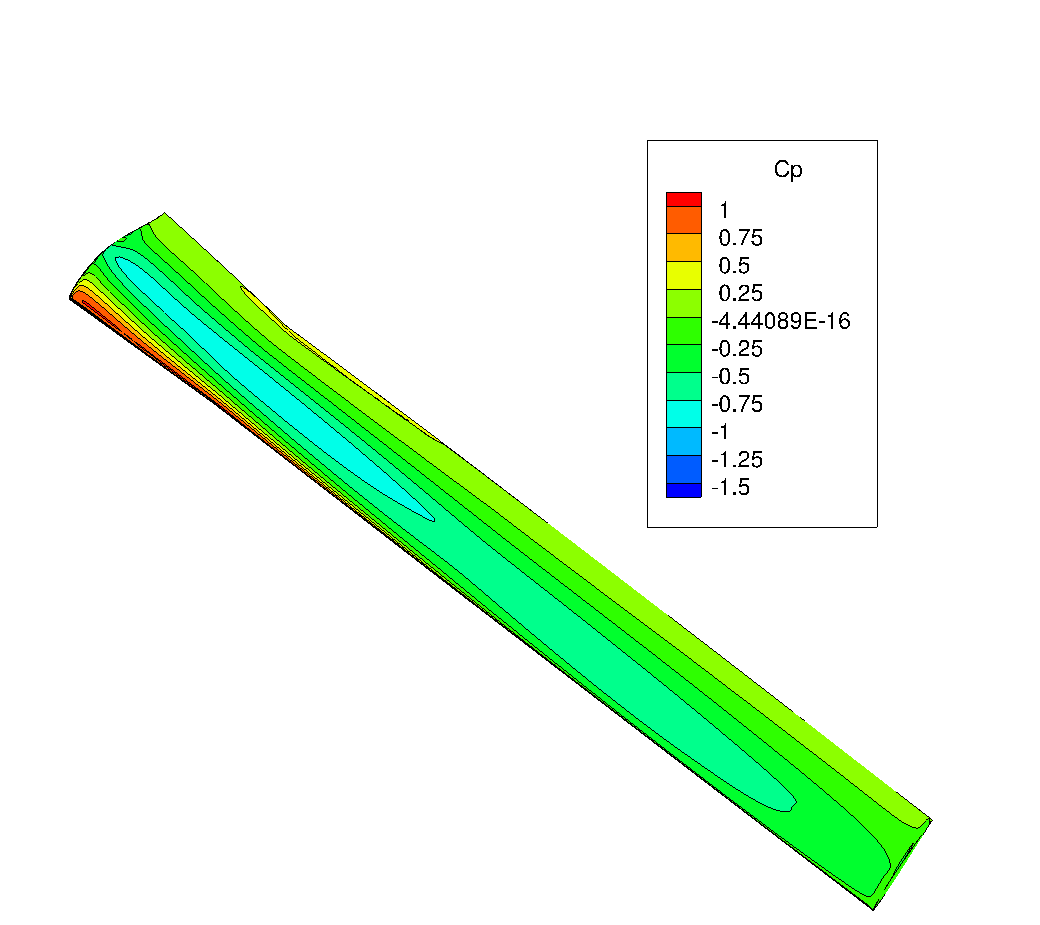}  
  \caption{Blade collective pitch angle $\theta_{75}=26\degree$}
  \label{fig:propeller_cp_pitch26}
\end{subfigure}
\begin{subfigure}{.5\textwidth}
  \centering
  \includegraphics[width=.8\linewidth]{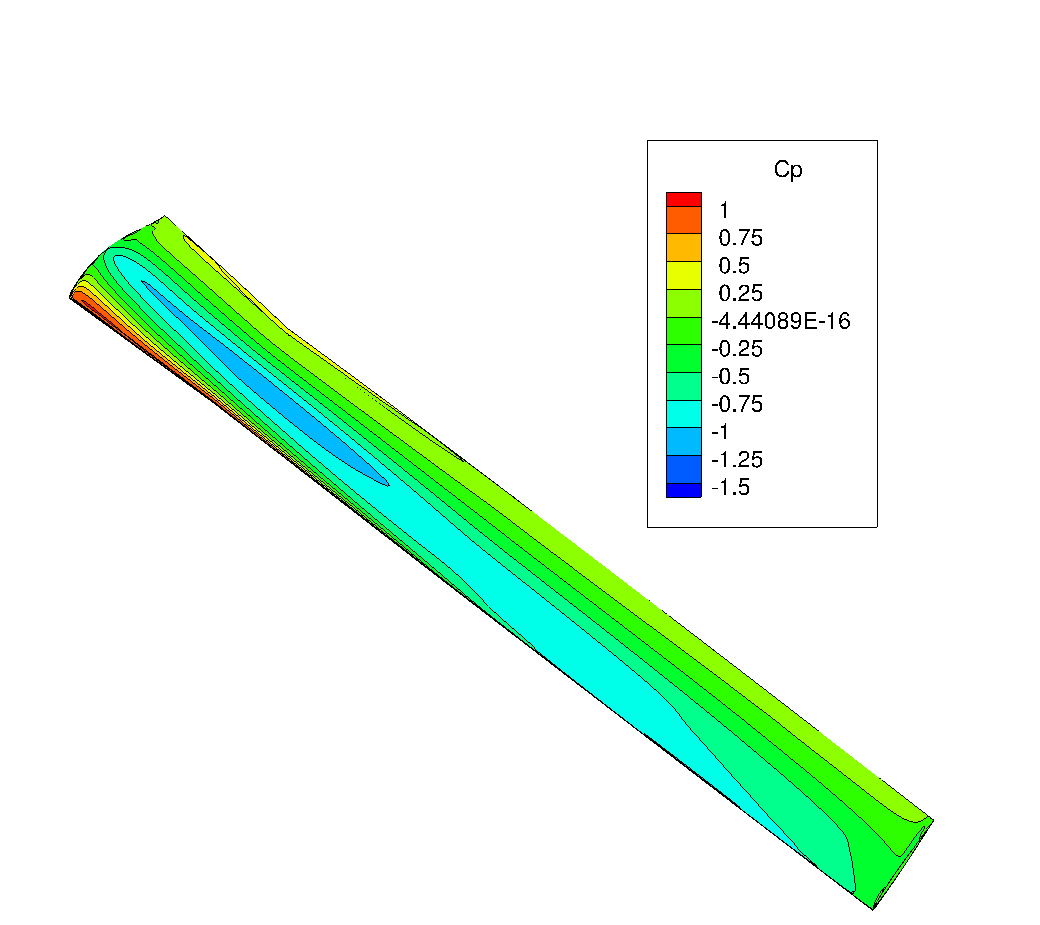}  
  \caption{Blade collective pitch angle $\theta_{75}=28.8\degree$}
  \label{fig:propeller_cp_pitch28.8}
\end{subfigure}
\caption{Contours of pressure coefficient for one XV-15 rotor blade in airplane mode.}
\label{fig:propeller_cp_contour}
\end{figure}

Figure \ref{fig:propeller_cp_contour} shows the predicted pressure coefficient distribution of the upper surface of a XV-15 blade at the advance ratio $\mu=0.337$. As the collective pitch angle increased, the span-wise blade loading becomes more uniform.

\subsubsection{Flowfield Details}
\begin{figure}[H]
\begin{subfigure}{.5\textwidth}
  \centering
  \includegraphics[width=.8\linewidth]{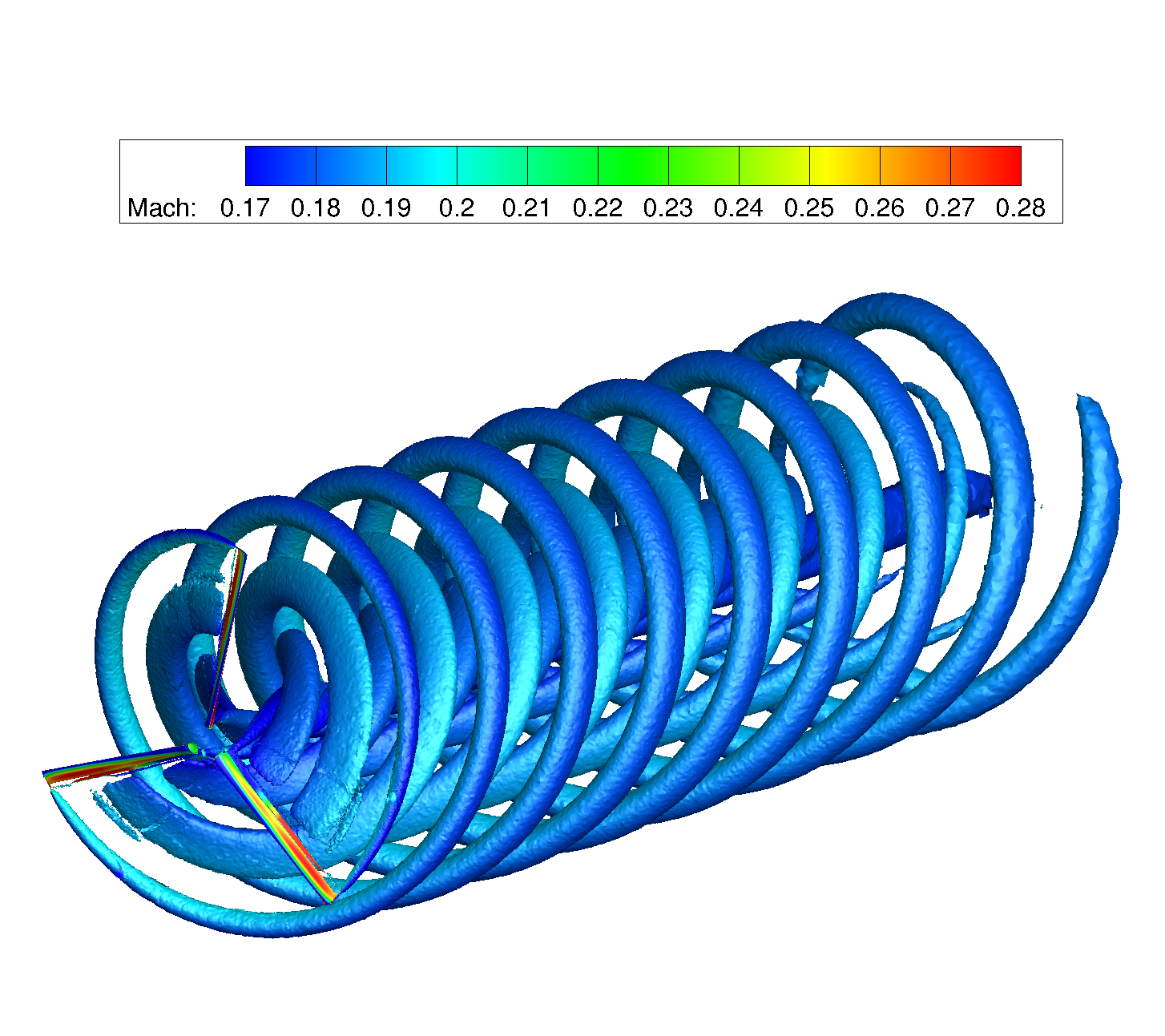}  
  \caption{Blade collective pitch angle $\theta_{75}=26\degree$}
  \label{fig:propeller_q_pitch26}
\end{subfigure}
\begin{subfigure}{.5\textwidth}
  \centering
  \includegraphics[width=.8\linewidth]{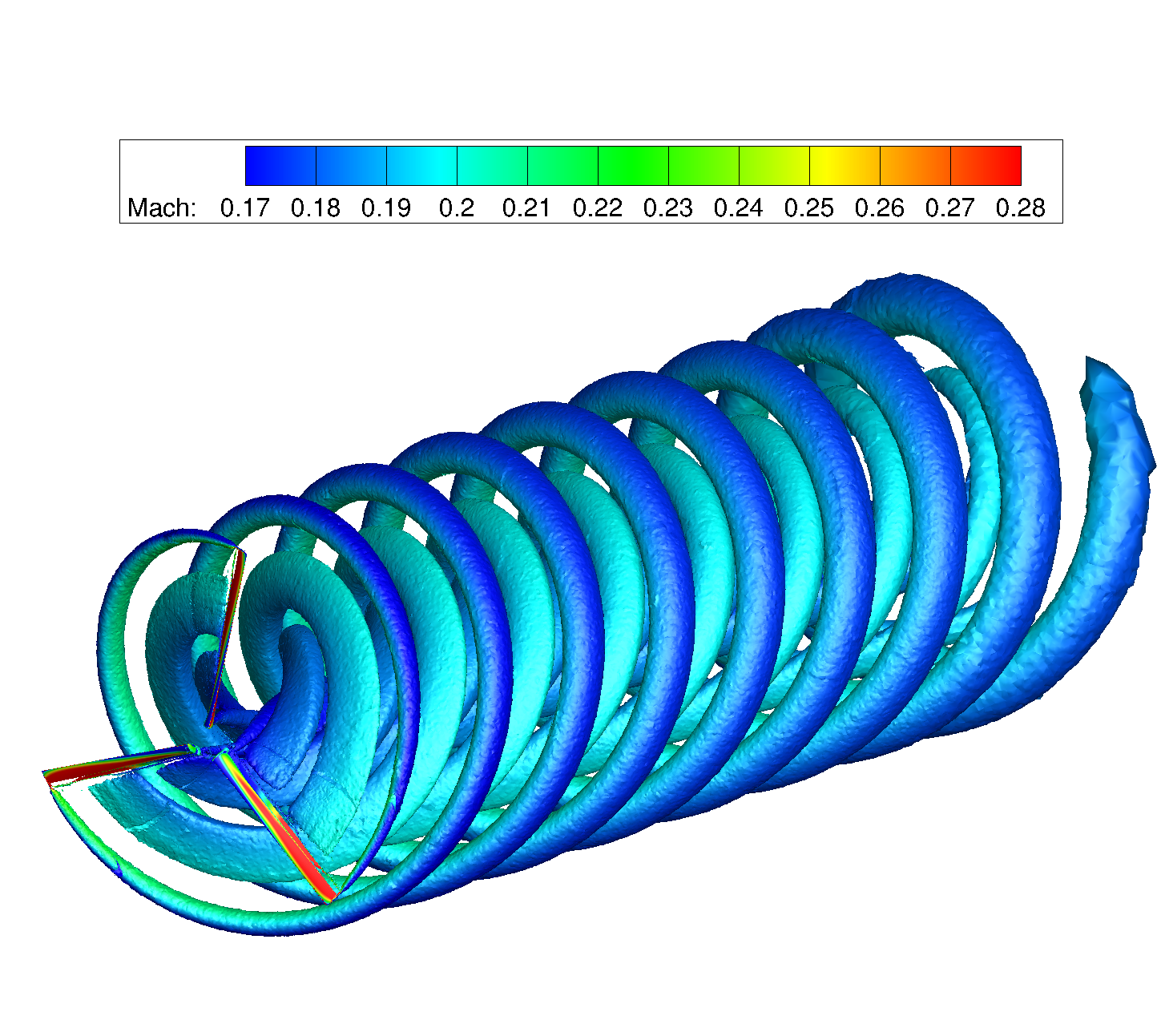}  
  \caption{Blade collective pitch angle $\theta_{75}=28.8\degree$}
  \label{fig:propeller_q_pitch28.8}
\end{subfigure}
\caption{Wake visualization of the airplane mode using Q-criterion shaded by contours of Mach at collective pitch of $\theta_{75}=26\degree$ (left) and $\theta_{75}=28.8\degree$ (right).}
\label{fig:propeller_qcriterion}
\end{figure}

Flowfield visualization of the rotor wake of two collective pitch angles in airplane mode, $\theta_{75}=26\degree$ and $28.8\degree$, using the Q criterion is given in Figure \ref{fig:propeller_qcriterion}. It can be seen that the vortices are stronger with a higher collective pitch angle. Moreover, because of the non-zero axial velocity of freestream, the blade vortex interaction is much less than the hovering flight, leading to simpler tip vortex trajectories.

\subsection{Forward Flight Helicopter Mode}
In the study of forward flight of helicopter mode, the shaft angle (angle of attack of the rotor tip-path-plane) varied from $-5\degree$ to $5\degree$ to simulate a wide range of both propulsive and descending forward flight conditions. The flow in this mode is much more complex because of the more intensive blade-vortex interactions. The advance ratio ($V_{\infty}/\Omega R$) is 0.17 mentioned in Table \ref{tab:flowCondition}. The detailed information of tested collective pitch angles for different shaft angle is listed in Table \ref{tab:twistAngleForForward}. The experimental data reported here were from Ref. \cite{Betzina2002RotorMode}, and they were already corrected to exclude the hub and tares effects.
\begin{table}[H]
\centering
\caption{\label{tab:twistAngleForForward}Collective pitch angles tested in forward flight of helicopter mode.}
\begin{tabularx}{0.8\textwidth} { 
  | >{\centering\arraybackslash}X 
  | >{\centering\arraybackslash}X|}
 \hline
 \textbf{shaft angle }$\bm{\alpha}$ & \textbf{collective pitch angles }$\bm{\theta_{75}}$ \\
 \hline
 -5\degree  & 4\degree, 5\degree, 6\degree, 7\degree, 8\degree, 9\degree, 10\degree\\
  \hline
  0\degree  & 3\degree, 5\degree, 6\degree, 7\degree, 8\degree, 9\degree\\
  \hline
  5\degree  & 2\degree, 3\degree, 4\degree, 5\degree, 6\degree, 7\degree\\
\hline
\end{tabularx}
\end{table}

\subsubsection{Overall Blade Loads and Flowfield details}
\begin{figure}[H]
\centering
\includegraphics[width=.8\linewidth]{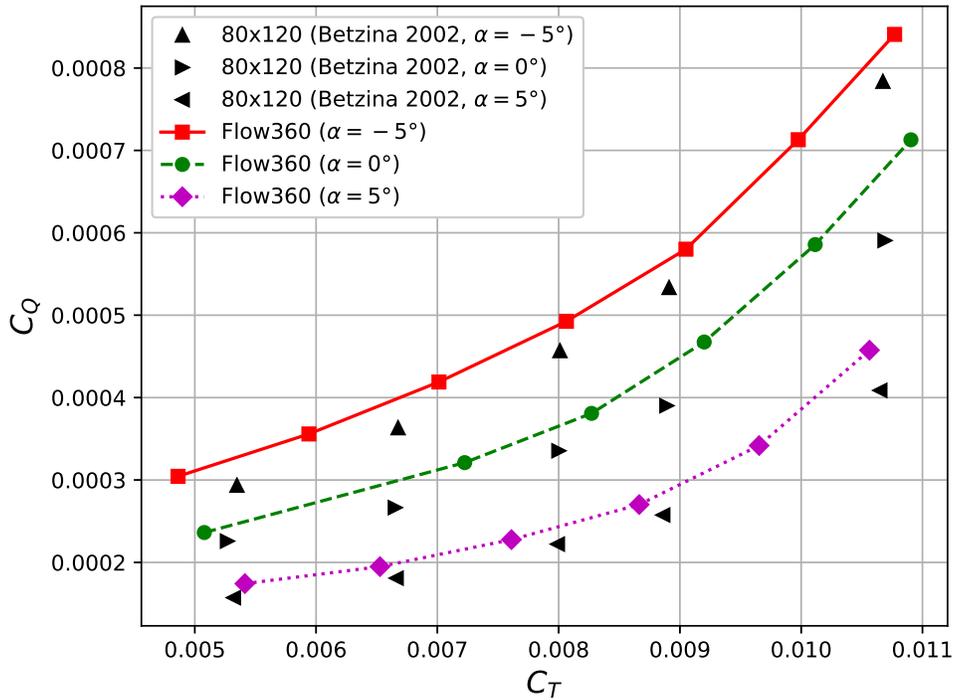}  
\caption{Comparison of measured and predicted torque coefficient versus thrust coefficient in forward flight helicopter mode with advance ratio $\mu=0.17$.}
\label{fig:forward_advanceRatio0.17_CTCQ}
\end{figure}

\begin{figure}[H]
\centering
\includegraphics[width=.8\linewidth]{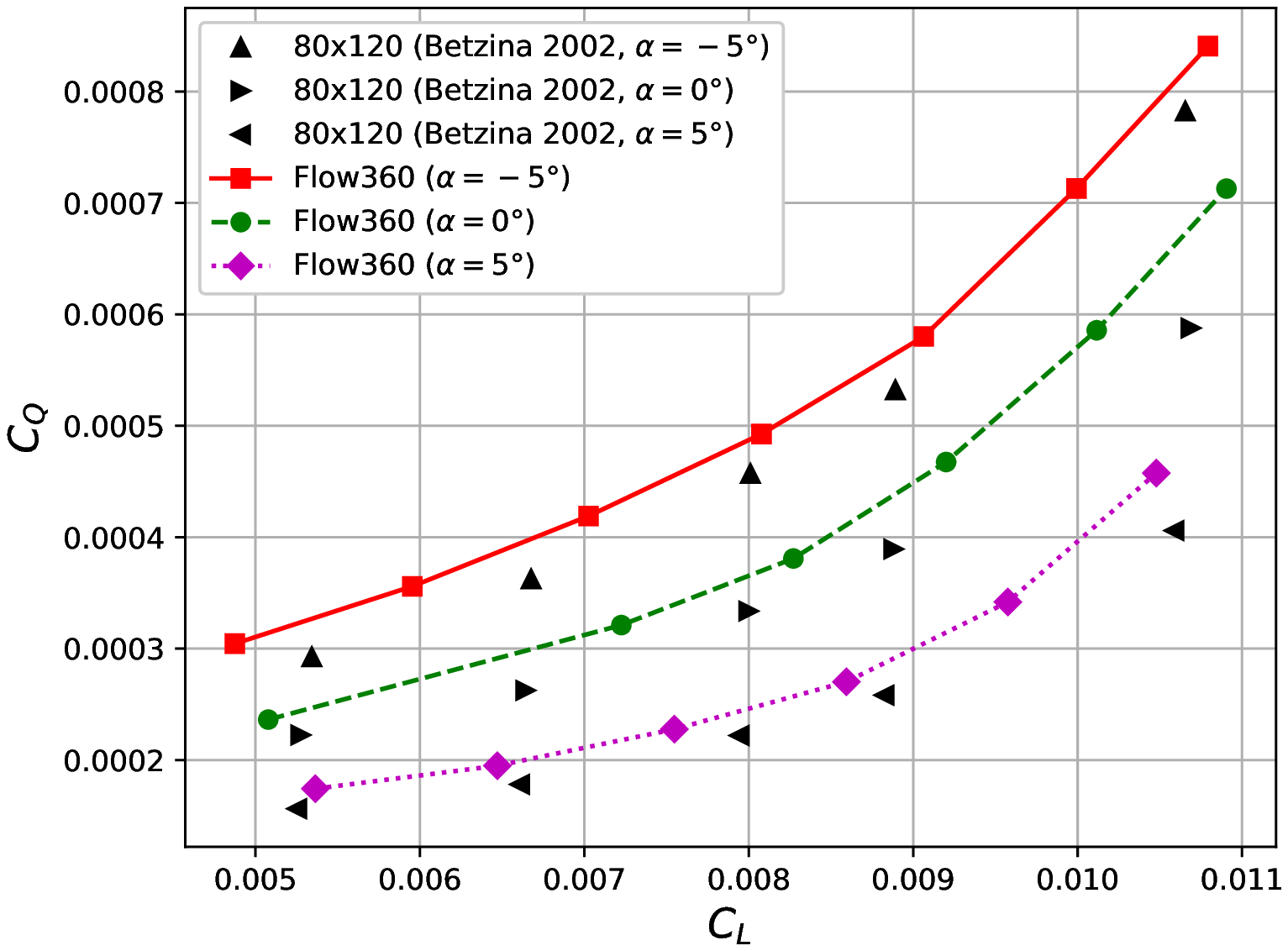}  
\caption{Comparison of measured and predicted torque coefficient versus lift coefficient in forward flight helicopter mode with advance ratio $\mu=0.17$.}
\label{fig:forward_advanceRatio0.17_CLCQ}
\end{figure}

\begin{figure}[H]
\begin{subfigure}{.5\textwidth}
  \centering
  \includegraphics[width=.8\linewidth]{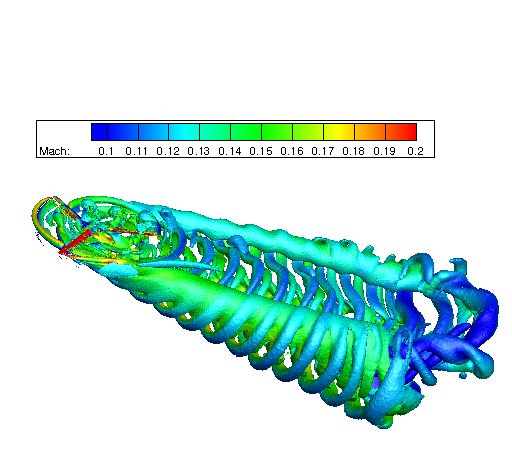}  
  \caption{Angle of attack of rotor disk $\alpha=-5\degree$}
  \label{fig:forward_q_alpha-5}
\end{subfigure}
\begin{subfigure}{.5\textwidth}
  \centering
  \includegraphics[width=.8\linewidth]{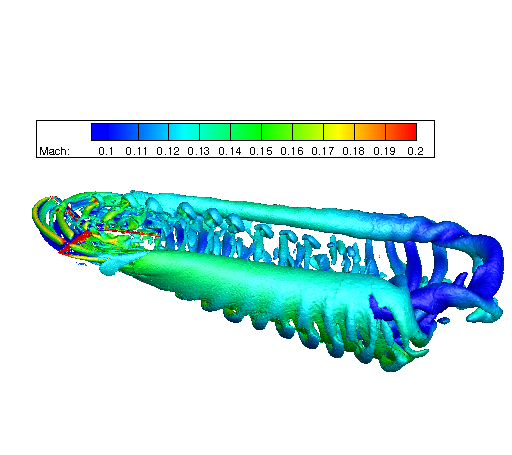}  
  \caption{Angle of attack of rotor disk $\alpha=5\degree$}
  \label{fig:forward_q_alpha5}
\end{subfigure}
\caption{Wake visualization of the helicopter forward flight mode using Q-criterion shaded by contours of Mach at propulsive $\alpha=-5\degree$ (left) and $\alpha=5\degree$ (right).}
\label{fig:forward_qcriterion}
\end{figure}

The relation between thrust and lift to torque is shown in Figure \ref{fig:forward_advanceRatio0.17_CTCQ} and Figure \ref{fig:forward_advanceRatio0.17_CLCQ}. The predicted rotor performance has $4\%-14\%$ relative error compared with the experimental data. Also, Figure \ref{fig:forward_qcriterion} shows the rotor wake in helicopter forward flight mode using the $Q$ criterion. It can be seen that the tip vortices have more interaction with the downstream blades. Because of the uncertainties in both simulations and measurements, a comparison between prediction and experimental data in greater physical detail should be done to find the reason for they discrepancy. However, due to lack of more precise experimental data, e.g. pressure distribution and skin friction, to verify and validate the Flow360 further, as a future work, an investigation on the impact of grid resolution and time integration needs to be done to study their impacts on capturing the underlying physics.

\section{Conclusions}
\label{section:conclusions}
The helicopter mode (hovering flight and forward flight) and airplane mode of a full-scale XV-15 tiltrotor are numerically studied using the Flow360 solver. All of the meshes involved are multi-block unstructured meshes, which are very efficient to create meshes with various collective pitch angles ($0\degree - 28.8\degree$ in the present study). The following conclusions are given the present study:
\begin{enumerate}
  \item The predicted integrated performance, e.g. $C_T$, $C_Q$, in hovering flight shows excellent agreement with the wind tunnel tests in~\cite{Felker1986PerformanceRotor, Light1997ResultsComplex, Betzina2002RotorMode}.
  \item The predicted blade sectional thrust and torque of hovering flight of helicopter mode capture the effect of tip vortices, which are close to the CFD data of HMB3.
  \item In hovering flight of helicopter mode, the predicted skin friction can match the measured data in most of turbulent flow regions, but a transition model is necessary to capture the natural transition process, which is a future area of research.
  \item In airplane mode with a medium advance ratio ($\mu=0.337$), the predicted integrated $C_T$ and $C_Q$ show good agreement with the experiment in \cite{Anon1971AdvancementResults}.
  \item In forward flight helicopter mode, the predicted $C_T$ and $C_Q$ matches the measured data fairly well. A grid resolution study and time integration study are useful to investigate the reasons of discrepancy.
\end{enumerate}

\bibliography{references}

\begin{thebibliography}{23}
\newcommand{\enquote}[1]{``#1''}
\providecommand{\natexlab}[1]{#1}
\providecommand{\url}[1]{\texttt{#1}}
\providecommand{\urlprefix}{URL }
\expandafter\ifx\csname urlstyle\endcsname\relax
  \providecommand{\doi}[1]{\discretionary{}{}{}https://doi.org/#1}\else
  \providecommand{\doi}[1]{\discretionary{}{}{}\urlstyle{rm}\url{https://doi.org/#1}}\fi

\bibitem[{Maisel et~al.(2000)Maisel, Giulianetti, and
  Dugan}]{Maisel2000TheFlight}
Maisel, M.~D., Giulianetti, D.~J., and Dugan, D.~C., \enquote{{The History of
  The XV-15 Tilt Rotor Research Aircraft: From Concept to Flight},} \emph{NASA
  Special Publication 4517}, 2000, p. 194.

\bibitem[{Potsdam et~al.(2004)Potsdam, Schaller, Rajagopalan, and
  Silva}]{Potsdam2004TiltFlight}
Potsdam, M.~A., Schaller, D.~F., Rajagopalan, R.~G., and Silva, M.~J.,
  \enquote{{Tilt rotor aeromechanics phenomena in low speed flight},} \emph{AHS
  International 4th Decennial Specialists' Conference on Aeromechanics}, 2004.

\bibitem[{Narducci et~al.(2009)Narducci, Jiang, Liu, and
  Clark}]{Narducci2009CFDInteractions}
Narducci, R., Jiang, F., Liu, J., and Clark, R., \enquote{{CFD modeling of
  tiltrotor shipboard aerodynamics with rotor wake interactions},}
  \emph{Collection of Technical Papers - AIAA Applied Aerodynamics Conference},
  2009.
\newblock \doi{10.2514/6.2009-3857}.

\bibitem[{Harris et~al.(2010)Harris, Scheidler, Hopkins, and
  Fortenbaugh}]{Harris2010InitialTiltrotor}
Harris, J.~C., Scheidler, P.~F., Hopkins, R., and Fortenbaugh, R.~L.,
  \enquote{{Initial power-off testing of the BA609 tiltrotor},} \emph{Annual
  Forum Proceedings - AHS International}, 2010.

\bibitem[{{Anon}(1971)}]{Anon1971AdvancementResults}
{Anon}, \enquote{{Advancement of Proprotor Technology Task 2 - Wind-Tunnel Test
  Results},} \emph{Nasa Cr 114363}, 1971.

\bibitem[{Weiberg and Maisel(1980)}]{Weiberg1980Wind-TunnelAircraft}
Weiberg, J.~A., and Maisel, M.~D., \emph{{Wind-Tunnel Tests of the XV-15 Tilt
  Rotor Aircraft}}, 1980.

\bibitem[{Felker et~al.(1986)Felker, Young, and
  Signor}]{Felker1986PerformanceRotor}
Felker, F., Young, L.~E., and Signor, D., \enquote{{Performance and Loads Data
  from a Hover Test of a Full-Scale Advanced Technology XV-15 Rotor},}
  \emph{Nasa Technical Memorandum 86-854}, , No. January 1986, 1986, p. 359.

\bibitem[{Light(1997)}]{Light1997ResultsComplex}
Light, J.~S., \enquote{{Results from an XV-15 rotor test in the national
  full-scale aerodynamics complex},} \emph{Annual Forum Proceedings - American
  Helicopter Society}, Vol.~1, No. May, 1997, pp. 231--239.

\bibitem[{Betzina(2002)}]{Betzina2002RotorMode}
Betzina, M.~D., \enquote{{Rotor Performance of an Isolated Full-Scale XV-15
  Tiltrotor in Helicopter Mode},} \emph{American Helicopter Society
  Aerodynamics, Acoustics, and Test and Evaluation Technical Specialists
  Meeting}, 2002, pp. 1--12.

\bibitem[{Wadcock et~al.(1999)Wadcock, Yamauchi, and
  Driver}]{Wadcock1999SkinRotor}
Wadcock, A.~J., Yamauchi, G.~K., and Driver, D.~M., \enquote{{Skin friction
  measurements on a hovering full-scale tilt rotor},} \emph{Journal of the
  American Helicopter Society}, Vol.~44, No.~4, 1999, pp. 312--319.
\newblock \doi{10.4050/JAHS.44.312}.

\bibitem[{Johnson(1980)}]{Johnson1980HelicopterTheory}
Johnson, W., \emph{{Helicopter Theory}}, Dover Publications, 1980.

\bibitem[{Landgrebe(1972)}]{Landgrebe1972WAKEPERFORMANCE.}
Landgrebe, A.~J., \enquote{{WAKE GEOMETRY OF A HOVERING HELICOPTER ROTOR AND
  ITS INFLUENCE ON ROTOR PERFORMANCE.}} \emph{Journal of the American
  Helicopter Society}, 1972.
\newblock \doi{10.4050/jahs.17.3}.

\bibitem[{Kocurek and Tangler(1977)}]{Kocurek1977PRESCRIBEDANALYSIS.}
Kocurek, J.~D., and Tangler, J.~L., \enquote{{PRESCRIBED WAKE LIFTING SURFACE
  HOVER PERFORMANCE ANALYSIS.}} \emph{Journal of the American Helicopter
  Society}, 1977.
\newblock \doi{10.4050/JAHS.22.24}.

\bibitem[{Schmitz et~al.(2009)Schmitz, Bhagwat, Moulton, Caradonna, and
  Chattot}]{Schmitz2009TheLoads}
Schmitz, S., Bhagwat, M., Moulton, M.~A., Caradonna, F.~X., and Chattot, J.~J.,
  \enquote{{The prediction and validation of hover performance and detailed
  blade Loads},} \emph{Journal of the American Helicopter Society}, Vol.~54,
  No.~3, 2009, pp. 0320041--03200412.
\newblock \doi{10.4050/JAHS.54.032004}.

\bibitem[{Yang et~al.(2002)Yang, Sankar, Smith, and
  Bauchau}]{Yang2002RecentFlight}
Yang, Z., Sankar, L.~N., Smith, M.~J., and Bauchau, O., \enquote{{Recent
  improvements to a hybrid method for rotors in forward flight},} \emph{Journal
  of Aircraft}, Vol.~39, No.~5, 2002, pp. 804--812.
\newblock \doi{10.2514/2.3000},
  \urlprefix\url{https://arc.aiaa.org/doi/abs/10.2514/2.3000}.

\bibitem[{Kaul and Ahmad(2011)}]{Kaul2011SkinOVERFLOW2}
Kaul, U.~K., and Ahmad, J., \enquote{{Skin friction predictions over a hovering
  tilt-rotor blade using OVERFLOW2},} \emph{29th AIAA Applied Aerodynamics
  Conference 2011}, , No. June, 2011.
\newblock \doi{10.2514/6.2011-3186}.

\bibitem[{Kaul(2012)}]{Kaul2012EffectFlows}
Kaul, U.~K., \enquote{{Effect of inflow boundary conditions on hovering
  tilt-rotor flows},} \emph{7th International Conference on Computational Fluid
  Dynamics, ICCFD 2012}, 2012.

\bibitem[{Yoon et~al.(2014)Yoon, Pulliam, and
  Chaderjian}]{Yoon2014SimulationsOverflow}
Yoon, S., Pulliam, T.~H., and Chaderjian, N.~M., \enquote{{Simulations of XV-15
  rotor flows in hover using overflow},} \emph{American Helicopter Society
  International - 5th Decennial AHS Aeromechanics Specialists' Conference 2014:
  Current Challenges and Future Directions in Rotorcraft Aeromechanics}, 2014,
  pp. 365--375.

\bibitem[{Sheng et~al.(2016)Sheng, Zhao, and
  Hill}]{Sheng2016InvestigationsCodes}
Sheng, C., Zhao, Q., and Hill, M., \enquote{{Investigations of XV-15 rotor
  hover performance and flow field using U2NCLE and HELIOS codes},} \emph{54th
  AIAA Aerospace Sciences Meeting}, Vol.~0, No. January, 2016, pp. 1--18.
\newblock \doi{10.2514/6.2016-0303}.

\bibitem[{Gates(2013)}]{Gates2013AerodynamicComputations}
Gates, S., \enquote{{Aerodynamic analysis of tiltrotors in hovering and
  propeller modes using advanced navier-stokes computations},} \emph{39th
  European Rotorcraft Forum 2013, ERF 2013}, 2013, pp. 106--109.

\bibitem[{Jimenez-Garcia et~al.(2017)Jimenez-Garcia, Barakos, and
  Gates}]{Jimenez-Garcia2017TiltrotorValidation}
Jimenez-Garcia, A., Barakos, G., and Gates, S., \enquote{{Tiltrotor CFD Part I
  - validation},} \emph{The Aeronautical Journal}, Vol. 121, No. 1239, 2017,
  pp. 577--610.
\newblock \doi{10.1017/aer.2017.17},
  \urlprefix\url{https://www.cambridge.org/core/product/identifier/S0001924017000173/type/journal_article}.

\bibitem[{Haimes and Dannenhoffer(2013)}]{Haimes2013TheGeometry}
Haimes, R., and Dannenhoffer, J., \enquote{{The Engineering Sketch Pad: A
  Solid-Modeling, Feature-Based, Web-Enabled System for Building Parametric
  Geometry},} \emph{21st AIAA Computational Fluid Dynamics Conference},
  American Institute of Aeronautics and Astronautics, Reston, Virginia, 2013.
\newblock \doi{10.2514/6.2013-3073},
  \urlprefix\url{http://arc.aiaa.org/doi/10.2514/6.2013-3073}.

\bibitem[{Spalart(2009)}]{Spalart2009Detached-eddySimulation}
Spalart, P.~R., \enquote{{Detached-eddy simulation},} , 2009.
\newblock \doi{10.1146/annurev.fluid.010908.165130}.

\end{thebibliography}
\end{document}